\RequirePackage{fix-cm}
\documentclass[smallextended]{svjour3}       % onecolumn (second format)
\smartqed  % flush right qed marks, e.g. at end of proof
\usepackage[pdftex]{graphicx}
\usepackage{amssymb}
\usepackage{amsmath}
\usepackage{theorem}
\theoremstyle{break}
\newtheorem{Theorem}{Theorem}%[section]

\def\Proof{\hfil\break{\bf Proof}\;\;\;\;}
\newtheorem{Proposition}{Proposition}%[section]
\newtheorem{Lemma}{Lemma}%[section]
\newtheorem{Corollary}{Corollary}%[section]
\newtheorem{Definition}{Definition}%[section]
\def\DIS{\displaystyle}
\def\hbreak{\vspace*{2mm}\hfill\break\noindent}
\def\e{\mbox{e}}

\def\Z{\mathbb{Z}}

\def\R{\mathbb{R}}

\begin{document}

\title{Rule 184 fuzzy cellular automaton as a mathematical model for traffic flow%\thanks{Grants or other notes
}

%\titlerunning{Rule 184 FCA as a traffic flow model}        % if too long for running head

\author{Kohei Higashi         \and
        Junkichi Satsuma \and Tetsuji Tokihiro %etc.
}


\institute{Kohei Higashi \at
             Graduate School of Mathematical Sciences, the University of Tokyo, 3-8-1 Komaba, Meguro-ku, Tokyo 153-8914, JAPAN   \\
              Tel.: +81-3-5465-7001\\
              Fax: +81-3-5465-7011\\
              \email{koheih@ms.u-tokyo.ac.jp}           %  \\
           \and
          Junkichi Satsuma \at
            Department of Mathematical Engineering, Faculty of Engineering, Musashino University, 3-3-3　Ariake, Koto-ku, Tokyo 202-8585, JAPAN\\
             Tel.: +81-3-6865-8093\\
              Fax:  +81-3-5465-7011\\
             \email{jsatsuma@musashino-u.ac.jp} 
            \and
            Tetsuji Tokihiro \at
            Graduate School of Mathematical Sciences, the University of Tokyo, 3-8-1 Komaba, Meguro-ku, Tokyo 153-8914, JAPAN   \\
              Tel.: +81-3-5465-7080\\
              Fax: +81-3-5465-7011\\
              \email{toki@ms.u-tokyo.ac.jp}              
}

\date{Received: date / Accepted: date}

\maketitle

\begin{abstract}
The rule 184 fuzzy cellular automaton is regarded  as a mathematical model of traffic flow because it contains the two fundamental traffic flow models, the rule 184 cellular automaton and the Burgers equation, as special cases. 
We show that the fundamental diagram (flux-density diagram) of this model consists of three parts: a free-flow part, a congestion part and a two-periodic part.
The two-periodic part, which may correspond to the synchronized mode region, is a two-dimensional area in the diagram, the boundary of which consists of the free-flow and the congestion parts.
We prove that any state in both the congestion and the two-periodic parts is stable, but is not asymptotically stable, while that in the free-flow part is unstable.
Transient behaviour of the model and  bottle-neck effects are also examined by numerical simulations. 
Furthermore, to investigate low or high density limit, we consider ultradiscrete limit of the model and 
show that any ultradiscrete state turns to a travelling wave state of velocity one in finite time steps for generic initial conditions.  
\end{abstract}

\section{Introduction}
\label{intro}
In modern society, efficient transportation system for goods and people is indispensable to the foundation of industry, hence it is necessary to analyse the system in detail. 
However, like the collective behaviour of cars on highways, it is generally very difficult to test and characterize them directly.
Therefore mathematical modelling of traffic flow has been performed since 1950s \cite{Lighthill-Whitham,Pipes} and various models have been constructed to reproduce empirical traffic flows. 
In the model, rigorous analysis and numerical simulations are used to clarify the basic properties such as traffic jams, density-flow diagram, bottleneck effect, etc. \cite{Elefteriadou,Nagatani}. 
The models are roughly classified into a macroscopic model and a microscopic model.
A macroscopic model is usually described by equations of macroscopic variables such as car density and average car velocity.
From analogy to flow of molecules in a liquid or a gas, the equations are often derived from the fundamental equations of fluid dynamics\cite{Musha-Higuchi}.
A simple equation for a macroscopic traffic model is the Burgers equation:
\begin{equation}\label{Intro_Burgers}
\frac{\partial \eta}{\partial \tau}(x,\tau)=2\eta(x,\tau)\frac{\partial \eta}{\partial x}(x,\tau)+\frac{\partial^2 \eta}{\partial x^2}(x,\tau).
\end{equation}
Here $\eta(x,\tau)$ $(x,\, \tau \in \R)$ is the normalized density of cars at position $x$ and time $\tau$ in appropriate units. 
Equation \eqref{Intro_Burgers} has a shock wave solution
\[
\eta(x,\tau)=\frac{k}{1+\e^{-kx-k^2\tau}}\qquad (k > 0)
\]
which shows that traffic jams propagate in the opposite direction of car movement, that is consistent with actual traffic flow.
While, in a microscopic model, a car is represented as a self-driven particle that moves in one direction, and its velocity changes depending on its position and/or speed relative to other particles\cite{Bando_etal,Sugiyama}.
A cellular automaton (CA) traffic model is a typical microscopic model \cite{Nagel-Schreckenberg}, in which the dynamics of cars is discretized in both time and space.  
Accordingly a state of traffic flow is expressed by an array of cells that take only finite number of states, and is updated in discrete time steps by a simple time evolution rule.
One of the most fundamental CA traffic model is the rule 184 CA in the elementary CAs (ECAs) defined by Wolfram\cite{Wolfram}.
An ECA is a one-dimensional two-states CA, and a state of a cell is updated with those of its adjacent two cells and itself at the previous time step.
Let us denote by  $u_n^t$ $\in \{0,1\}$ the state of $n$th cell at time step $t$ ($n,\,t \in \Z$).
The updating rule for the rule 184 CA is given as
\begin{equation}
u_n^{t+1}=\left\{
\begin{array}{cc}
1&\;\; (u_{n-1}^t=1, u_n^t=0)\;\;
\mbox{or}\;\; ( u_{n}^t=u_{n+1}^t=1)\\
0&\;\;\mbox{otherwise}
\end{array}
\right.
\label{Intro_eq1}
\end{equation} 
As a traffic model, we suppose that a single-lane road is divided into pieces of an appropriate inter-vehicle distance and number them in the direction of traffic flow. If there is a car in the $n$th section at time step $t$, we put $u_n^t=1$, otherwise  $u_n^t=0$.  
Equation \eqref{Intro_eq1} means that a car will move to the next section if and only if it is not occupied by the car in front.
Although the rule 184 is very simple, it can reproduce the congestion phenomenon; the fundamental diagram, which gives the relation
of the flux (the average velocity of cars multiplied by the density of them) to the density, shows sharp transition from free-flow region to congested region as the density of cars increases. 
It is noted that the rule 184 CA can be regarded as the ultradiscrete analogue of the Burgers equation\cite{Nishinari-Takahashi}.    

In this article, we investigate a traffic model\cite{Higashi_etal}, which may be considered as a macroscopic extension of the rule 184 CA.  
We derive the model in the next section and show that it includes the rule 184 CA in a special case and that its continuous limit gives the Burgers equation. 
In section 3, all the stationary states are obtained and classified for cyclic boundary conditions, and we present the fundamental diagrams of the model and prove that this model has stable two-dimensional region of so called synchronized mode\cite{Kerner-Rehborn,Lubashevski_etal}  as well as free-flow region and congested region. 
For open boundary conditions, we show analytic expression of the steady states and discuss the bottleneck effect with numerical simulations.
In section 4, we perform ultradiscrete analysis of the model to examine travelling waves in low density, and prove that any initial state turns to a travelling wave state in finite time steps.
Section 5 is devoted to the concluding remarks. 

%
%
%,
%
\section{Rule 184 fuzzy CA}
\label{sec:1}
%\begin{figure}[bht]
%\centering
%\includegraphics[width=.8\textwidth]{./fig/figure1-1.eps}
%\caption{Schematic figure for the present traffic model.}
%\end{figure}
%
%\subsection{Model}
\begin{figure}[bht]
\centering
\includegraphics[width=.8\textwidth]{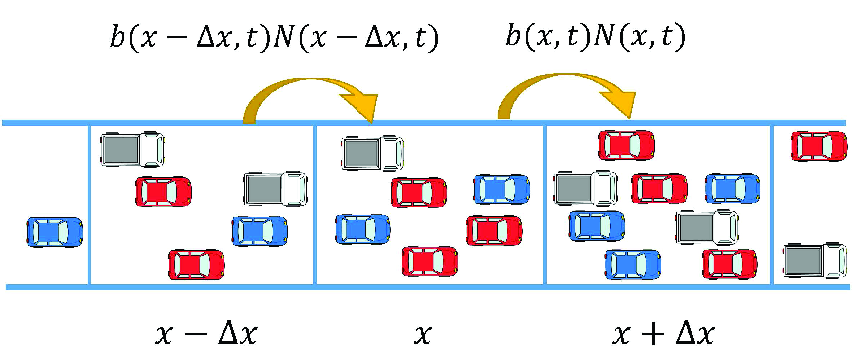}
\caption{Schematic figure for the present traffic model.}
\label{fig:1}       % Give a unique label
\end{figure}
We consider a multi-lane road and divide it into one-dimensional array of sections by appropriate distance $\Delta x$ (Fig~\ref{fig:1}).
Let $N(x,\tau)$ be a number of cars in the section $[x, x+\Delta x]$ at time $\tau$.
Because of equation of continuity, we have
\begin{equation}\label{sec1_cont}
\frac{N(x,\tau+\Delta \tau)-N(x,t)}{\Delta \tau}=\frac{J(x,\tau)-J(x+\Delta x,\tau)}{\Delta x},
\end{equation}
where $J(x,\tau)$ is the flux of cars at position $x$ and time $\tau$, and, roughly speaking, is equal to the average velocity of cars multiplied by the density of cars.
The average velocity in general depends on the density of cars and is a decreasing function of the density.
The function is sometimes called $k-v$ relation, and is approximately a linearly decreasing function\cite{Greenshield,Ni}.
Hence we may assume
\begin{equation}\label{sec1_kv}
J(x,\tau) \propto  N(x-\Delta x, \tau)\left(1-\frac{N(x,\tau)}{N_{\max}} \right),
\end{equation}
where $N_{\max}$ is the maximum number of cars in a section.

Let us normalize the equation \eqref{sec1_cont}.
We define 
\begin{equation}\label{sec1_rho_N}
\rho_n^t:=\frac{N(n\Delta x, t \Delta\tau)}{N_{\max}} \qquad (0 \le \rho_n^t \le 1),
\end{equation}
and, accordingly, we put
\[
j_n^t := \frac{\Delta \tau}{N_{\max}\Delta x}J(n\Delta x, t \Delta \tau).
\]
From \eqref{sec1_kv}, the normalized flux $j_n^t$ may be given as 
\begin{equation}\label{sect1_flux}
j_n^t=\rho_{n-1}^t(1-\rho_n^t).
\end{equation}
Therefore we have  
\begin{equation}\label{sec1_FCA184}
\rho_n^{t+1}=\rho_{n-1}^t(1-\rho_n^t) + \rho_n^t\rho_{n+1}^t.
\end{equation}
Note that, from \eqref{sec1_FCA184}, if $0 \le \rho_{n-1}^t,\, \rho_{n+1}^t \le 1$,  then
\[
0 \le \rho_n^{t+1} \le  (1-\rho_n^t) + \rho_n^t=1.
\]  
Thus, for any initial state $\{ \rho_n^{t=0}\}$ $(\,{}^\forall n,\  0 \le \rho_n^0 \le 1)$, it holds that ${}^\forall n,{}^\forall t,\; 0 \le \rho_n^t \le 1$.
Furthermore, if ${}^\forall n$, $\rho_n^0\in \{0,1\}$, then ${}^\forall n,{}^\forall t,$ $\rho_n^t \in \{0,1\}$ and
\[
\rho_n^{t+1}=\left\{
\begin{array}{cc}
1&\;\; (\rho_{n-1}^t=1, \rho_n^t=0)\;\;
\mbox{or}\;\; ( \rho_{n}^t=\rho_{n+1}^t=1)\\
0&\;\;\mbox{otherwise}
\end{array}
\right.,
\]
which is the same time evolution rule as that of the rule 184 CA \eqref{Intro_eq1}.
Since \eqref{sec1_FCA184} is a discrete dynamical system in both time and space, and its dependent variables take continuous values in $[0,1]$, we can consider \eqref{sec1_FCA184} as a continuous CA.  
A continuous CA the updating rule of which is given by \textit{fuzzification} of the original Boolian CA is called a fuzzy CA (FCA)\cite{Cattaneo_etal}.
Hence, we call the dynamical system described by \eqref{sec1_FCA184} the rule 184 FCA, or FCA184 in abbreviation.

To consider a continuous limit of \eqref{sec1_FCA184} with respect to its independent variables, we put \eqref{sec1_rho_N} into \eqref{sec1_FCA184},
\[
\frac{N(x,\tau+\Delta \tau)}{N_{\max}}=\left[\frac{N(x-\Delta x, \tau)}{N_{\max}}\left(1-\frac{N(x,\tau)}{N_{\max}}\right) + \frac{N(x,\tau)N(x+\Delta x,\tau)}{N_{\max}^2}    \right].
\]
As a small fluctuation around $\frac{N_{\max}}{2}$, we introduce $\eta(x,\tau)$ by the following formula:
\[
\frac{N(x,\tau)}{N_{\max}}=\frac{1}{2}\left(1+\Delta x \eta (x,\tau)\right).
\]
By taking Taylor series expansion
\begin{align*}
\frac{N(x,\tau+\Delta \tau)}{N_{\max}}&=\frac{1}{2}\left(1+\Delta x \eta (x,\tau+\Delta \tau)\right)\\
&=\frac{1}{2}\left[1+\Delta x \left\{ \eta (x,\tau)+\Delta \tau \frac{\partial \eta }{\partial \tau}(x,\tau) +O(\Delta \tau^2)\right\}\right],
\\%\end{align*}
%\begin{align*}
\frac{N(x\pm \Delta x,\tau)}{N_{\max}}&=\frac{1}{2}\left(1+\Delta x \eta (x\pm \Delta x,\tau)\right)\\
&=\frac{1}{2}\left[1+\Delta x \left\{ \eta (x,\tau)\pm \Delta x \frac{\partial \eta }{\partial x}(x,\tau) + \frac{\Delta x^2}{2} \frac{\partial^2 \eta }{\partial x^2}(x,\tau) +O(\Delta x^3)\right\}\right],
\end{align*}
we have 
\[
\frac{\partial \eta}{\partial \tau}(x,\tau)+O(\Delta \tau)=\frac{\Delta x^2}{\Delta \tau}\eta(x,\tau)\frac{\partial \eta}{\partial x}(x,\tau) +\frac{\Delta x^2}{2\Delta \tau}\frac{\partial^2 \eta}{\partial x^2}(x,\tau) +O\left(\frac{\Delta x^3}{\Delta \tau} \right).
\]
Thus, to take a limit $\Delta x \rightarrow 0$, $\Delta \tau \rightarrow 0$ with constraint $\frac{\Delta x^2}{2\Delta \tau}=1$, we obtain
%\[
%\frac{\partial \eta}{\partial \tau}(x,\tau)=2\eta(x,\tau)\frac{\partial \eta}{\partial x}(x,\tau) +\frac{\partial^2 \eta}{\partial x^2}(x,\tau),
%\]
%which is 
the Burgers equation \eqref{Intro_Burgers}.
Thus we find that the FCA184 \eqref{sec1_FCA184} contains the rule 184 CA as a special case, and that the density fluctuation of FCA184 around $\rho_n^t=\frac{1}{2}$ is described by the Burgers equation.
\section{Stationary states and fundamental diagram of FCA184}
\label{sec:2}
Statistical properties of traffic flow are empirically investigated by the fundamental diagram, the diagram which displays the relation between density and flux, and it is one of the most important objects which characterize a traffic model.   
To establish the fundamental diagram of FCA184,  we adopt a periodic boundary condition in which the total number of cars does not change:
\begin{equation}\label{sect2_boundary1}
\rho_n^t=\rho_{n+N}^t, \qquad (N \in \Z_{>0}).
\end{equation}
An interesting feature of stationary and asymptotically stationary states is that they depend on the parity of $N$.
We shall discuss other boundary conditions later in this section. 

The average density $\langle \rho\rangle$, which is a constant in time, is defined as
\begin{equation}\label{sect2_density}
\langle \rho\rangle:=\frac{1}{N}\sum_{n=1}^N\rho_n^t,
\end{equation}
and the average flux $J^t$ at time $t$ is given from \eqref{sect1_flux} by
\begin{equation}\label{sect2_flux}
J^t:=\frac{1}{N}\sum_{n=1}^N\rho_{n-1}^t(1-\rho_n^t).
\end{equation}
Note that $\rho_0^t=\rho_N^t$.
For a state $\{\rho_n^t\}$, we have a pair $(\langle \rho \rangle,\, J^t)$, and 
the fundamental diagram is a two-dimensional plot of these pairs.
We define a (multi-valued) function $Q(s)$ which takes the values of $J^t$ for a given density $\langle \rho \rangle = s$. The fundamental diagram is exhibited in the two dimensional $s$-$Q$ plane.
An important fundamental diagram is that for stationary states.
Here a stationary state is the state which realizes at $t \rightarrow \infty$ for an initial state.
More precisely, we define it as follows.
\begin{Definition}\label{def_stationary}
For any $\epsilon >0$, if there exist integers $T$ and $L$ such that 
\[
{}^\forall n,\,{}^\forall t,\;  |\rho_n^t-\rho_{n+L}^{t+T}|<\epsilon, 
\]
the solution $\{\rho_n^t\}$ of \eqref{sec1_FCA184} is called a stationary state.  
\end{Definition} 
The definition \ref{def_stationary} implies that a quasi-periodic solution is also a stationary state, however, as is shown later, FCA 184 with a periodic boundary condition does not have a quasi-periodic solution.
\subsection{Stationary states and fundamental diagram of FCA 184}
\label{sect2_1}
First we consider the case $N=2M$ ($M \in \Z_{>0}$).
It is readily seen that there are two types of stationary states.
\hbreak
\textbf{Uniform state} \par 
There is a trivial state ${}^\forall n,{}^\forall t,\, \rho_n^t=s$ ($0 \le s  \le 1$).
In this case, $J^t=s(1-s)$ and the fundamental diagram is  given by the function
\begin{equation}\label{sect2_uniform}
Q(s)=s(1-s) \quad (0 \le s \le 1).
\end{equation}
\hbreak
\textbf{Travelling wave state}　\par
Let $\rho_n^t=\sigma_{n-t}^{t}$ and find a solution of FCA184 which satisfies $\sigma_j^t=\sigma_j$.
Since
\[
\sigma_{j-1}=\sigma_{j-1}(1-\sigma_j)+\sigma_j\sigma_{j+1},
\]
we have
\[
\sigma_j(\sigma_{j+1}-\sigma_{j-1})=0\quad \sigma_j\sigma_{j+1}=\sigma_{j-1}\sigma_j.
\]
Thus we obtain the following solutions.
\hbreak
$\;\;$\textbf{1) two-periodic solution}\; If ${}^\forall j,\ \sigma_j \ne 0$, then, we find
\[
{}^\forall j \quad \sigma_{j-1}=\sigma_{j+1}.
\]
Hence, $\sigma_{2m-1}=\alpha,\quad \sigma_{2m}=\beta$  \quad ($1 \le m \le M$).
If $\alpha=\beta$, we have a uniform state.	
Hence a uniform state is a special case of the two-periodic state.

The flux is caluculated as
\[
J^t=\frac{1}{2}\left(\alpha(1-\beta)+\beta(1-\alpha)\right).
\]
Since the average density $\langle \rho \rangle$ is equal to  $s=\frac{\alpha+\beta}{2}$, 
by putting $\alpha=s+c,\;\beta=s-c$, the function $Q(s)$ is determined as
\begin{equation}
Q(s)=s(1-s)+c^2\qquad (|c| \le \min(s,1-s)). 	
\end{equation}
\hbreak
$\;\;$\textbf{2) free flow solution}\; In the case ${}^\exists i_0,\ \sigma_{i_0}=0$, 
${}^\forall j,\ \sigma_j\sigma_{j+1}=\sigma_{j-1}\sigma_j$. Thus we find 
\[
\sigma_1\sigma_2=\sigma_2\sigma_3 = \cdots =
\sigma_N\sigma_1=0.
\]
Therefore a solution must satisfy either $\sigma_j=0$ or $\sigma_{j+1}=0$ for an arbitrary $j$. 
For example, with a set of $M$ values $\{\gamma_1,\gamma_2,...,\gamma_M\}$ $(0 \le \gamma_m \le 1)$, 
\[
\sigma_{2m}=0,\quad \sigma_{2m+1}=\gamma_m \qquad (m=1,2,...,M)
\]
is one of such solutions.
A free flow solution shows a travelling wave going forward with velocity $1$. 

The average flux is given as
\[
J:=\frac{1}{2M}\sum_{j=1}^N\sigma_{j-1}(1-\sigma_j).
\]
Noticing the fact that $\sigma_j \ne 0$ $\rightarrow$ $\sigma_{j-1}=0$, we have
\[
J=\frac{1}{2M}\sum_{\sigma_j \ne 0}\sigma_j=\frac{1}{2M}\sum_{j=1}^N\sigma_j.
\]
Since the average density $s$ is equal to $\frac{1}{2M}\sum_{j=1}^N\sigma_j$, we find
\begin{equation}\label{sect2_freeflow}
Q(s)=s \qquad \left(0 \le s \le \frac{1}{2}\right).
\end{equation}
Note that $J^t \le \frac{1}{2}$. （There is no solution for $s>\frac{1}{2}$）.
\hbreak
$\;\;$\textbf{3)  anti-free flow solution}\; 
By putting $q_n^t=1-\rho_n^t$, \eqref{sec1_FCA184} turns into 
\begin{equation}\label{sect2_eq1_r}
q_n^{t+1}=q_{n+1}^t(1-q_n^t)+q_n^tq_{n-1}^t \quad(1 \le n \le N).
\end{equation}
Thus we see that, if $\rho_n^t$ is a solution to \eqref{sec1_FCA184}, then $q_n^t=\rho_{-n}^t$ is a solution to \eqref{sect2_eq1_r}.
Accordingly, by putting $j=n+t$, $r_j:=q_n^t$ satisfies
\[
r_{j+1}=r_{j+1}(1-r_j)+r_jr_{j-1} \ \rightarrow r_jr_{j-1}=r_{j+1}r_j.
\]
A solution corresponding to the two-periodic solution is also a two-periodic solution, but, there is another kind of solutions which satisfy 
either $r_j=0$ or  $r_{j+1}=0$ for any $j$.
This condition implies that either $\rho_n^t=1$ or $\rho_{n+1}^t=1$ for arbitrary $n$, and $\rho_n^t=\rho_{n+t}^0$.
This solution, an anti-free flow solution, shows a travelling wave which goes backward with velocity $1$.

Since 
\[
\frac{1}{N}\sum_{j=1}^N r_j=\frac{1}{N}\sum_{n=1}^N (1-\rho_n^t)=1-s,
\]
the average flux is calculated as
\begin{align*}
J&=\frac{1}{N}\sum_{n=1}^{N}\rho_{n-1}^t(1-\rho_n^t)\\
&=\frac{1}{N}\sum_{n=1}^{N}q_{n}^t(1-q_{n+1}^t)\\
&=\frac{1}{N}\sum_{j=1}^{N}r_j(1-r_{j+1})\\
&=\frac{1}{N}\sum_{j=1}^{N}r_j=1-s.
\end{align*}
Here we use the fact $r_jr_{j+1}=0$.
Thus we find 
\begin{equation}
Q(s)=1-s \qquad  \left(\frac{1}{2}\le s \le 1\right).
\end{equation}

In case of $N=2M+1$ ($M \in \Z_{>0}$), by repeating the similar arguments as above, we find that there exist uniform states, 
free flow states and anti-free flow states, but no two-periodic state exists because of the periodic boundary condition.
The uniform states have the same function $Q(s)$ as in the case of $N$ even;
\begin{equation}
Q(s)=s(1-s) \qquad (0 \le s \le 1).
\end{equation}  
For the free flow states, the function $Q(s)$ is given as
\begin{equation}\label{sect2_odd_free}
Q(s)=s \qquad \left(0 \le s \le \frac{M}{N}\right),
\end{equation}
and for the anti-free flow states
\begin{equation}\label{sect2_odd_anti}
Q(s)=1-s \qquad \left(\frac{M+1}{N} \le s \le 1\right).
\end{equation}

\begin{figure}[bht]
\centering
\includegraphics[width=.8\textwidth]{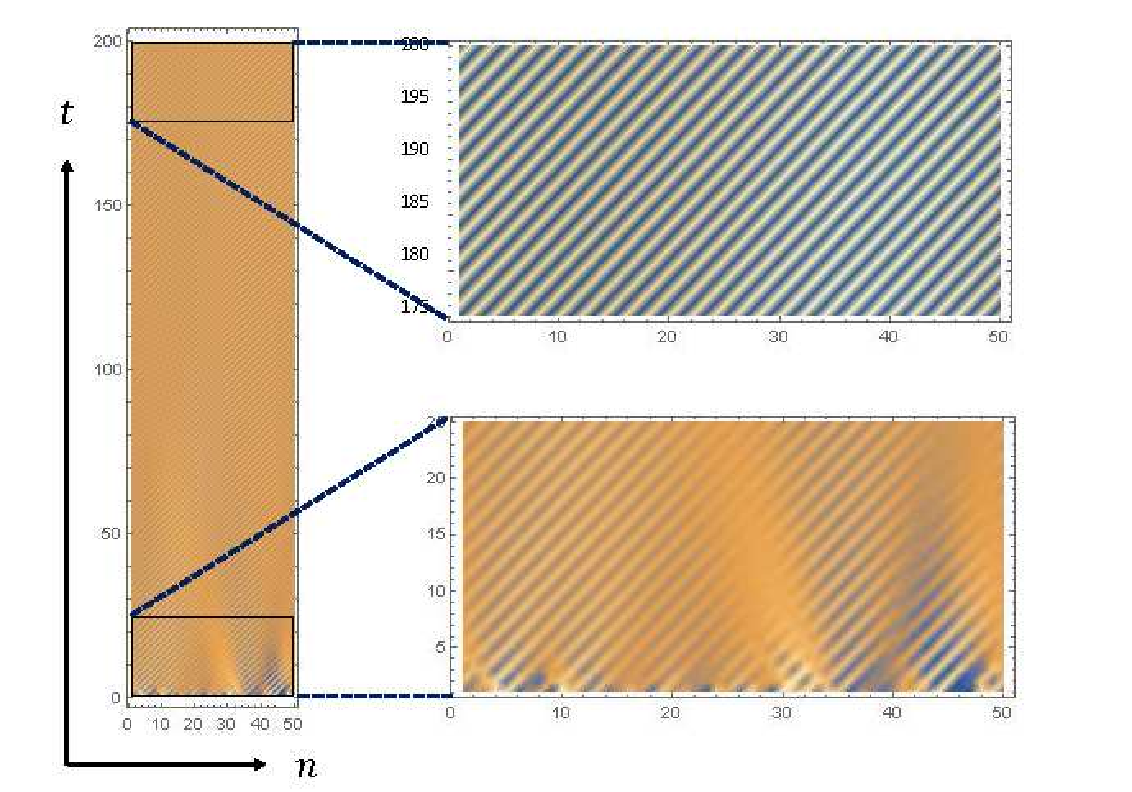}
\caption{A transient behaviour of FCA184 in the case of even $N$ ($N=50$). 
The lighter the colour, the greater the value. The state soon converges into a two-periodic state.}
\label{fig:4}       % Give a unique label
\end{figure}

\begin{figure}[bht]
\centering
\includegraphics[width=.8\textwidth]{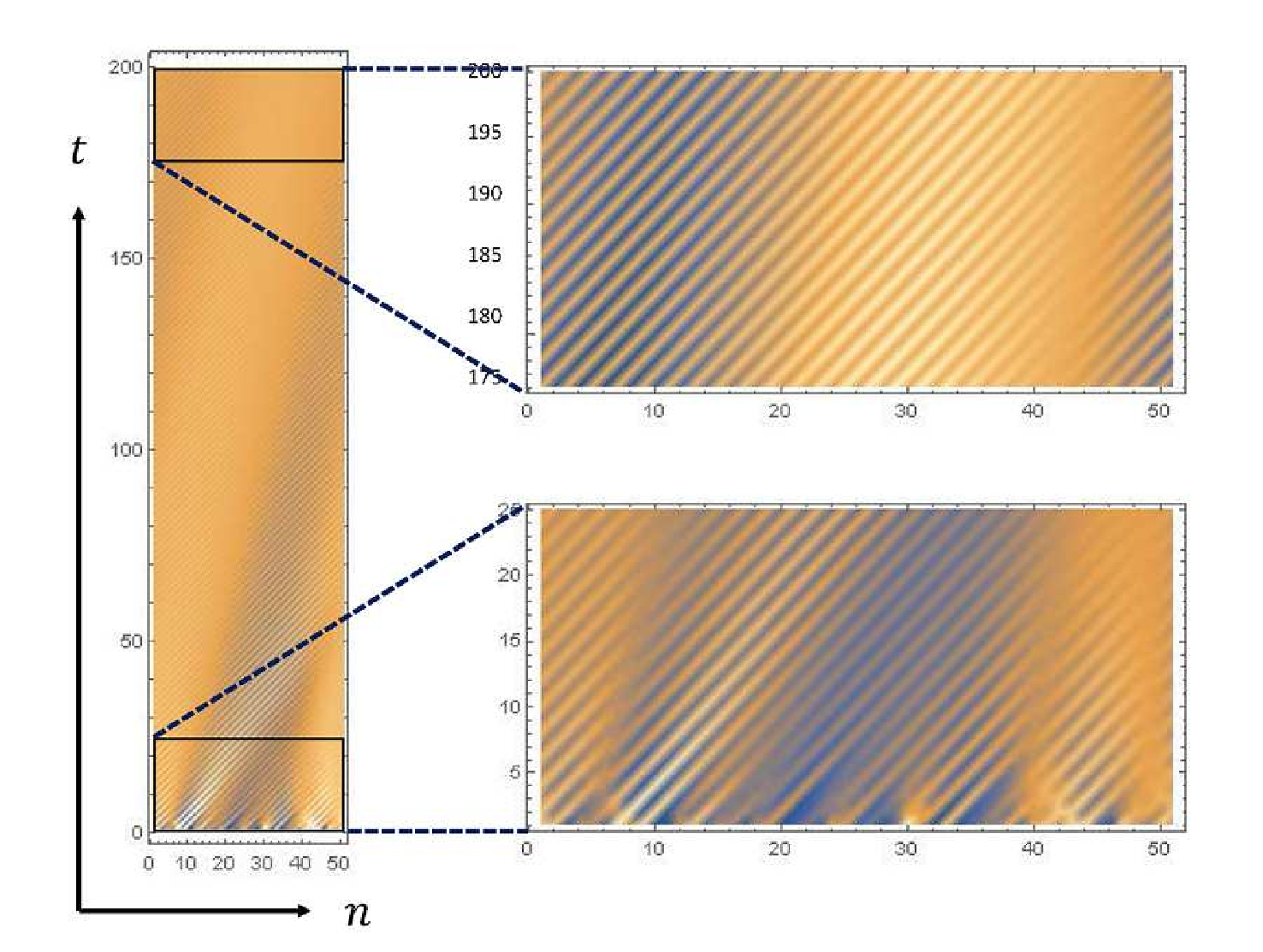}
\caption{A transient behavior of FCA184 in the case of odd $N$ ($N=51$). 
The lighter the colour, the greater the value. The state shows a feature of a two-periodic state for a long period of time.}
\label{fig:5}       % Give a unique label
\end{figure}

As will be proved in the next subsection (Theorem~\ref{sec2.2.theorem}), stationary solutions of FCA 184 are all that were listed above. 
Thus, in summary, we have the following Theorem. 
\begin{Theorem}\label{sec2_Th_1}
When the number of the sections $N$ is even, the fundamental diagram of the present traffic model for stationary states is the two-dimensional region:
\begin{equation}\label{fundamental_diagram_even}
s(1-s) \le Q \le \min[s,1-s] \quad (0 \le s \le 1). 
\end{equation}
While that for odd $N$ ($N=2M+1)$ is the one-dimensional boundary of the region \eqref{fundamental_diagram_even} which consists of the three parts;
\begin{equation}\label{fundamental_diagram_odd}
\left\{
\begin{array}{ll}
Q=s(1-s) &\;\; (0 \le s \le 1 )\\
Q=s &\;\; (0 \le s \le\frac{M}{N}  )\\
Q=1-s &\;\; (\frac{M+1}{N} \le s \le 1)
\end{array}
.
\right.
\end{equation}
\end{Theorem}
\begin{figure}[bht]
\centering
\includegraphics[width=.6\textwidth]{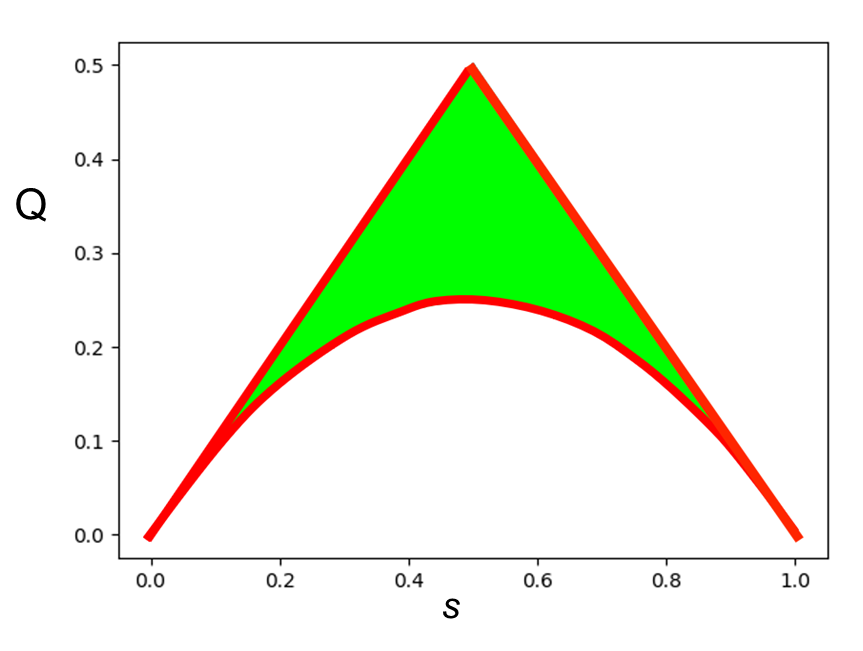}
\caption{The fundamental diagram of FCA184 in stationary states. When the total number of sites $N$ is even, the fundamental diagram is the two-dimensional area ($s(1-s) \le Q \le \min[s,1-s] $), while $N$ is odd, it is the one-dimensional boundary of this area.}
\label{fig:2}       % Give a unique label
\end{figure}%
One may think it strange that the fundamental diagram depends on the parity of the total number of sites.  
In fact, the features of traffic flow will not be affected by a boundary condition for $N$ $\rightarrow$ $\infty$.
Figure~\ref{fig:4} shows an example of time evolution of FCA184 in case of even $N$ with a periodic boundary condition.
The initial value of each site is generated randomly, and we see that the state soon converges into a two-periodic state.
While Fig.~\ref{fig:5} shows the case of odd $N$. 
Although it will converges to a uniform state, the state shows a feature of a two-periodic state over a long period of time.  
It is experimentally observed that three qualitative different types of traffic exists in a multi-lane traffic: free traffic flow, synchronized traffic flow, and traffic jams\cite{Kerner-Rehborn}.
Figure~\ref{fig:5} suggests that the two-periodic part for odd $N$ is metastable, that is, a state in this part is not strictly stable but is long-lived.
Hence we presume that these states correspond to the metastable synchronize modes which are observed in different models\cite{Lubashevski_etal,Tomer_etal}.
%
%
%
%
%\newpage
\subsection{Stability of stationary states}
The time evolution rule of FCA 184 is regarded as a weighted average rule defined by Betel and Flocchini \cite{Betel-Flocchini}.
Asymptotic properties of  FCAs with weighted average rules have been investigated in Ref.~\cite{Betel-Flocchini}, and the following 
proposition was proved.
\begin{Proposition}[Betel-Flocchini: Theorem3.9]\label{Prop_Betel}
If it holds that ${}^\forall n,\, 0<\rho_n^{t=0} <1$, the state of FCA 184 with a periodic boundary condition asymptotically converges to a two-periodic state when $N$ is even,
and to a uniform state when $N$ is odd.
\end{Proposition}
In order to make the present article self-contained, we give a proof  of  the Proposition \ref{Prop_Betel} for the case where $N$ is even.
If $N$ is odd, proof is performed in a similar way and is easier.

Prior to the proof of the proposition, we prepare a Lemma \ref{sec2.2.lemma1}.
Let us define $v_n^t$ by $\rho_n^t=:v_{n-t}^t$.
Because of the cyclic boundary condition, $v_{n+2M}^t=v_n^t$, and 
\[
v_n^{t+1}=(1-v_{n+1}^t)v_n^t+v_{n+1}^tv_{n+2}^t.
\]
Introducing $x_i^t$ and $y_i^t$ as 
\[
x_i^t:=v_{2i-1}^t,\;\; y_i^t:=v_{2i}^t \qquad (i=1,2,...,M),
\]
we find that $x_{i+M}^t=x_i^t$, $y_{i+M}^t=y_i^t$, and 
\begin{subequations}
\begin{align}
x_i^{t+1}&=(1-y_i^t)x_i^t+y_i^tx_{i+1}^t\label{2cycle_eqa}\\
y_i^{t+1}&=(1-x_{i+1}^t)y_i^t+x_{i+1}^ty_{i+1}^t \label{2cycle_eqb}.
\end{align}
\end{subequations}
\begin{Lemma}\label{sec2.2.lemma1}
The following inequalities hold.
\begin{subequations}
\begin{align}
\min_{i}\left[x_i^{t+1}\right] &\ge \min_{i}\left[x_i^{t}\right] \label{eq2a}\\
\max_i\left[x_i^{t+1}\right] &\le \max_i\left[x_i^t\right]  \label{eq2b}\\ 
\min_{i}\left[y_i^{t+1}\right] &\ge \min_{i}\left[y_i^{t}\right]  \label{eq2c}\\
\max_i\left[y_i^{t+1}\right] &\le \max_i\left[y_i^t\right]  \label{eq2d}
\end{align}
\end{subequations}
\end{Lemma}
\Proof
In \eqref{2cycle_eqa}, the inequality $0 \le y_i^t \le 1$ implies 
\[
\min[x_i^t,x_{i+1}^t] \le x_i^{t+1} \le \max[x_i^t,x_{i+1}^t].
\]
Hence we have
\[
\min_i\left[ \min[x_i^t,x_{i+1}^t]  \right]\le \min_i\left[x_i^{t+1}\right]. 
\]
The left hand side of the above inequality is equal to $\DIS \min_i\left[ x_i^t \right]$, and the inequality \eqref{eq2a} holds.
The other inequalities \eqref{eq2b} -- \eqref{eq2d} are proved similarly.
\qed
\vspace*{5pt}

Since, from the Lemma~\ref{sec2.2.lemma1}，the sequence $\DIS \left(\min_i\left[x_i^t\right]\right)_{t=0}^\infty$ is a monotonically increasing sequence with respect to $t$ and is bounded above, it converges to a certain real number. 
The other sequences like $\DIS \left(\max_i\left[x_i^t\right]\right)_{t=0}^\infty$ also converge, and we have
\begin{subequations}
\begin{align}
\lim_{t \to \infty} \min_{i}\left[x_i^{t}\right]=:\alpha_1 \label{eq3a}\\
\lim_{t \to \infty} \max_{i}\left[x_i^{t}\right]=:\alpha_2 \label{eq3b}\\ 
\lim_{t \to \infty} \min_{i}\left[y_i^{t}\right]=:\beta_1  \label{eq3c}\\
\lim_{t \to \infty} \max_{i}\left[y_i^{t}\right]=:\beta_2  \label{eq3d}
\end{align}
\end{subequations}
where $0 \le \alpha_1,\alpha_2,\beta_1,\beta_2 \le 1$.

The following Lemma is readily obtained from \eqref{2cycle_eqa} and \eqref{2cycle_eqb}.
\begin{Lemma}\label{sect2.2.lemma2}
If all the initial values are in the open interval $(0,1)$, inequalities
\[
0<x_i^t<1,\quad 0<y_i^t <1
\]
hold for any $i$ and any $t$. 
\end{Lemma}

%Here we define a set $\{c_i(n;t)\}$ as follows.
\begin{Definition}\label{def_c_int}
Fix the integers $t$ and $n$. 
By using \eqref{2cycle_eqa}, $c_n(i,t+s)$ ($s=0,1,2,...,M-2$), which are polynomials of \{$y_i^\tau\}$, are
defined successively as follows.
\begin{align*}
x_n^{t+M-1}&=(1-y_n^{t+M-2})x_n^{t+M-2}+y_n^{t+M-2}x_{n+1}^{t+M-2}\\
&:=c_n(n;t+M-2)x_n^{t+M-2}+c_{n+1}(n;t+M-2)x_{n+1}^{t+M-2}\\
&=(1-y_n^{t+M-2})\left\{ (1-y_n^{t+M-3})x_n^{t+M-3}+y_n^{t+M-3}x_{n+1}^{t+M-3}   \right\} \\
&\qquad +y_n^{t+M-2}\left\{ (1-y_{n+1}^{t+M-3})x_{n+1}^{t+M-3}+y_{n+1}^{t+M-3}x_{n+2}^{t+M-3}    \right\}\\
&=(1-y_n^{t+M-2})(1-y_n^{t+M-3})x_n^{t+M-3}+\left\{(1-y_n^{t+M-2})y_n^{t+M-3}\right. \\
&\qquad \left.+ y_n^{t+M-2} (1-y_{n+1}^{t+M-3}) \right\}x_{n+1}^{t+M-3}
+y_n^{t+M-2}y_{n+1}^{t+M-3}x_{n+2}^{t+M-3}\\
&=:c_n(n;t+M-3)x_n^{t+M-3}+c_{n+1}(n;t+M-3)x_{n+1}^{t+M-3}\\
&\qquad \qquad +c_{n+2}(n;t+M-3)x_{n+2}^{t+M-3}\\
&=\cdots \\
&=:\sum_{i=n}^{n+M-1}c_i(n;t) x_i^t
\end{align*}
For $k<n$, we define $c_k(i,t+s) := 0$.

We also define $\delta$ as
\[
\delta:=\min_i\left[ \min[y_i^0,1-y_i^0]\right]=\min\left[\min_i[y_i^0], 1-\max_i[y_i^0]\right].
\]
\end{Definition}
Note that, from Lemma~\ref{sect2.2.lemma2}, \eqref{eq2c} and \eqref{eq2d}, the following inequality holds for any $t$ and any $n$.
\begin{equation}\label{sec2.2.eta_ineq}
0<\delta \le y_n^t,\qquad 0<\delta \le 1-y_n^t.
\end{equation}

Furthermore, for $s=M-k-1$, we have 
\begin{subequations}
\begin{align}
&x_n^{t+M-1}\nonumber \\
&=\sum_{i=n}^{n+M-k-1}c_i(n;t+M-k) x_i^{t+M-k} \label{l2eq1}\\
&=\sum_{i=n}^{n+M-k-1}c_i(n;t+M-k)\nonumber \\
&\qquad\qquad \times \left\{(1-y_i^{t+M-k-1})x_i^{t+M-k-1}+y_i^{t+M-k-1}x_{i+1}^{t+M-k-1}\right\}\label{l2eq2}\\
&=\sum_{i=n}^{n+M-k} \left\{c_i(n;t+M-k)(1-y_i^{t+M-k-1})\right. \nonumber \\
&\qquad\qquad\qquad \left. +c_{i-1}(n;t+M-k)y_{i-1}^{t+M-k-1}\right\}x_i^{t+M-k-1}\label{l2eq3}\\
&=\sum_{i=n}^{n+M-k} c_i(n;t+M-k-1)  x_i^{t+M-k-1}\label{l2eq4}.
\end{align}
\end{subequations}
%where we used that $c_{n-1}(n;t+M-k) = 0$. 
From \eqref{l2eq3} and \eqref{l2eq4}, we find a recurrence relation:
\begin{align}
&c_i(n;t+M-k-1)\nonumber \\
&=c_i(n;t+M-k)(1-y_i^{t+M-k-1})+c_{i-1}(n;t+M-k)y_{i-1}^{t+M-k-1}.\label{l2eq5}
\end{align}
\begin{Lemma}\label{sect2.2.lemma3}
\begin{equation}\label{2.2.lem3eq}
\sum_{i=n}^{n+M-s-1}c_i(n;t+s)=1,\qquad \delta^{M-1-s} \le c_i(n;t+s)
\end{equation}
\end{Lemma}
\Proof
We prove by induction of $s$, starting from $s=M-2$ and going downwards.

Equation \eqref{2.2.lem3eq} clearly holds for $s=M-2$.

Suppose that it holds up to $s=M-k$ ($2 \le k \le M-1$).

If ${}^\forall i$, $x_i^{t+M-k-1}=1$, then ${}^\forall i$, $x_i^{t+M-k}=1$.
Hence, by taking ${}^\forall i$, $x_i^{t+M-k-1}=1$ in \eqref{l2eq1},  \eqref{l2eq4} leads to  
\[
\sum_{i=n}^{n+M-k-1}c_i(n;t+M-k) =\sum_{i=n}^{n+M-k} c_i(n;t+M-k-1) =1.
\]
Noticing the inequality $c_i(n;t+M-k) \ge \delta^{k-1}$, from \eqref{sec2.2.eta_ineq} and \eqref{l2eq5},
\[
c_i(n;t+M-k-1)\ge c_i(n;t+M-k)(1-y_i^t) \ge \delta^{k}.
\]
Thus \eqref{2.2.lem3eq} is true for $s=M-k-1$. 
Therefore, by the induction hypothesis, \eqref{2.2.lem3eq} holds for $0 \le s \le M-2$.
\qed\vspace*{5pt}

%Now we prove Proposition \ref{Prop_Betel}.
\hbreak
\textbf{Proof of Proposition \ref{Prop_Betel}}
\par\vspace*{5pt}
We prove by contradiction.

Since $0<\alpha_1 \le \alpha_2<1$, we suppose that $\alpha_1<\alpha_2$.
From \eqref{eq3a}, \eqref{eq3b} and the monotonicity of the sequences, for any $\epsilon >0$, there exists $t_\epsilon$ such that if $t \ge t_\epsilon$, then 
\[
\alpha_1-\epsilon  < \min_i[x_n^t] \le \alpha_1,\quad \alpha_2 \le \max_i[x_n^t] < \alpha_2 + \epsilon.
\]
Let $d:=\alpha_2-\alpha_1$ and choose $\epsilon$ so that  the inequality $0<\epsilon < d \delta^{M-1}$ holds.
For $t \ge t_\epsilon$, let $n$ and $m$ be the integers that satisfy
\[
x_n^{t+M-1}=\max_i[x_i^{t+M-1}],\quad x_m^t=\min_i[x_i^t].
\]
Then, from Lemma~\ref{sect2.2.lemma3},
\begin{align*}
x_n^{t+M-1}&=\sum_{i=n}^{n+M-1}c_i(n;t) x_i^t\\
&=c_m(n;t)x_m^t+\sum_{i=n,\,i\ne m}^{n+M-1}c_i(n;t) x_i^t\\
&<c_m(n;t)\alpha_1+\sum_{i=n,\,i\ne m}^{n+M-1}c_i(n;t) (\alpha_2+\epsilon)\\
&=c_m(n;t)\alpha_1+(1-c_m(n;t))(\alpha_2+\epsilon)\\
&<\alpha_2+\epsilon-c_m(n;t)(\alpha_2-\alpha_1)\\
&\le \alpha_2+\epsilon-d\delta^{M-1}\\
&<\alpha_2.
\end{align*}
Therefore $x_n^{t+M-1}<\alpha_2$.
However, by definition,
\[
x_n^{t+M-1}=\max_i\left[x_i^{t+M-1} \right] \ge \alpha_2,
\]
which is a contradiction.
Thus we find that $\DIS \alpha_1=\alpha_2$.

The equality $\beta_1=\beta_2$ can be proved in the same way and the Proposition~\ref{Prop_Betel} is proved. 
\qed\vspace*{10pt}

%%%%%%%%%%%%%%%%%%%%%%%%%%%%%%%%%%%%%%%%%%%%%%%%%
%%%%%%%%%%%%%%%%%%%%%%%%%%%%%%%%%%%%%%%%%%%%%%%%%
Now we will prove that all the stationary solutions are the uniform solutions and the travelling wave solutions listed in the previous subsection~\ref{sect2_1}. 
When the initial values contain neither $0$ nor $1$, Proposition~\ref{Prop_Betel} means that there is no stationary solution other than the uniform and the two-periodic solutions.
Hence we examine stationary states with $0$s and/or $1$s.
In such a state, there are four kinds of patterns which have a series of $0$s between non-zero values as
\begin{align*}
&P^{(00)}:=\{P_k^{(00)}\}_{k=1}^{N-1}, \qquad P_k^{(00)}:=\ast \underbrace{0\cdots 0}_{k} \ast \\
&P^{(10)}:=\{P_k^{(10)}\}_{k=1}^{N-2}, \qquad P_k^{(10)}:=1 \underbrace{0\cdots 0}_{k} \ast \\
&P^{(10)}:=\{P_k^{(01)}\}_{k=1}^{N-2}, \qquad P_k^{(01)}:=\ast \underbrace{0\cdots 0}_{k} 1 \\
&P^{(11)}:=\{P_k^{(11)}\}_{k=1}^{N-1}, \qquad P_k^{(11)}:=1 \underbrace{0\cdots 0}_{k} 1. 
\end{align*}
Here $\ast$ indicates any value other than $0$ and $1$. 

For a given state,  we define the number of the above patterns at time $t$ as 
\begin{equation}\label{numberofzeros}
l_k^{(ij)}(t):=\mbox{number of the pattern $P_k^{(ij)}$ at time $t$}\quad (i,j \in \{0,1\}).
\end{equation}
%Similarly we define for a sequence of $1$s as
%\begin{align*}
%Q^{(00)}:=\{Q_k^{(00)}\}, \qquad Q_k^{(00)}:=0 \underbrace{1\cdots 1}_{k} 0 \\
%Q^{(10)}:=\{Q_k^{(10)}\}, \qquad Q_k^{(10)}:=\ast \underbrace{1\cdots 1}_{k} 0 \\
%Q^{(10)}:=\{Q_k^{(01)}\}, \qquad Q_k^{(01)}:=0 \underbrace{1\cdots 1}_{k} \ast\\
%Q^{(11)}:=\{Q_k^{(11)}\}, \qquad Q_k^{(11)}:=\ast \underbrace{1\cdots 1}_{k} \ast \\
%\end{align*}
%
%

We consider time evolution of FCA 184 by introducing a new variable $v_{n}^t=\rho_{n+t}^t$:
\begin{equation}\label{sect2.2.proofeq1}
v_n^{t+1}=(1-v_{n+1}^t)v_n^t+v_{n+1}^tv_{n+2}^t
\end{equation}
with a periodic boundary condition 
$v_{n+N}^t=v_n^t$.
\begin{Proposition}\label{sect2.2.proofprop1}
Let  $M_0^t$ be the number of $0$s at time $t$, and $M_1^t$ be that of $1$s.
Then it holds that
\begin{equation}\label{01num_eq} 
M_0^{t+1} \le M_0^t,\qquad M_1^{t+1} \le M_1^t.
\end{equation}
\end{Proposition}
Note that 
\[
M_0^t=\sum_{k=1}^{N-1}\; k\left(l_k^{(00)}(t)+l_k^{(10)}(t)+l_k^{(01)}(t)+l_k^{(11)}(t)\right).
\]
\Proof
For a state at time step $t$, $v_n^{t+1}=0$ is achieved in the three cases:
\[
\begin{array}{cccc}
&v_n^t&v_{n+1}^t&v_{n+2}^t\\
\mbox{i)}&0&a&0\\
\mbox{ii)}&a&1&0\\
\mbox{iii)}&0&0&a
\end{array}
\]
where $a$ is arbitrary.
%
% Similarly $v_n^{t+1}=1$ is achieved in the cases:
%\[
%\begin{array}{cccc}
%&v_n^t&v_{n+1}^t&v_{n+2}^t\\
%\mbox{i)'}&1&a&1\\
%\mbox{ii)'}&1&0&a\\
%\mbox{iii)'}&a&1&1
%\end{array}\] 
Hence we can count the number of $0$s at $t+1$ as follows. 
\begin{itemize}
\item For $k=1$, the patterns $10*$ and $101$ contribute to the case ii), and the numbers of them are $l_1^{(10)}(t)$ and $l_1^{(11)}(t)$ respectively.
\item For $k=2$,  the pattern $*00*$ contributes to the case iii)，$100*$ to ii) and iii), $*001$ to iii), $1001$ to ii) and iii). The numbers are  $l_2^{(00)}(t),\, 2l_2^{(10)}(t),\,l_2^{(01)}(t)$, and $2 l_2^{(11)}(t)$ respectively.
\item For $k \ge 3$,  each pattern contains $k-2$ sequences of $000$ which contributes to i). Therefore, each of four patterns generates $(k-1)l_k^{(00)}(t),\, kl_k^{(10)}(t),\,(k-1)l_k^{(01)}(t)$ and $kl_k^{(11)}(t)$ $0$s respectively.
\item Besides these, patterns $0\ast 0$ ($\ast \ne 0,1$) and $010$ can exist in the state at the boundary of those four patterns. 
Let the number of the sequence $0\ast 0$ be $b(t)$.
This sequence appears in the three patterns:
\[
\begin{array}{ccc}
\ast 0\cdots 0 \ast 0\cdots 0 \ast, \qquad  &1 0\cdots 0 \ast 0\cdots 0 \ast, \qquad & \ast 0\cdots 0 \ast 0\cdots 0 1 \\
(P_k^{(00)}P_{k'}^{(00)})        &    (P_k^{(10)}P_{k'}^{(00)})        &   (P_k^{(10)}P_{k'}^{(01)})
\end{array}
\]
The total number of the first and second pattens is less than the number of the sequences $\ast 0\cdots 0 \ast$, that of the third pattern is less than the number of  the pattern $\ast 0\cdots 0 1$.
Thus
\[
b(t) \le \sum_{k=1}^{N-1} l_k^{(00)}(t)+l_k^{(01)}(t).
\]
The sequence $010$ appears in the following patterns:
\[
\begin{array}{ccc}
\ast 0\cdots 0 1 0\cdots 0 \ast, \qquad  &1 0\cdots 0 1 0\cdots 0 \ast, \qquad &\ast 0\cdots 0 1 0\cdots 0 1 \\
(P_k^{(01)}P_{k'}^{(10)})        &    (P_k^{(11)}P_{k'}^{(10)})        &   (P_k^{(01)}P_{k'}^{(11)})
\end{array}
\]
But contribution of these patterns was already counted as the case ii).
\end{itemize}
From the above consideration, 
\begin{align*}
M_0^{t+1}&=\left( \sum_{k=1}^{N-1} (k-1)l_k^{(00)}(t) + kl_k^{(10)}(t)+(k-1)l_k^{(01)}(t) + kl_k^{(11)}(t) \right) +b(t) \\
&\le \sum_{k=1}^{N-1}\; k\left(l_k^{(00)}(t)+l_k^{(10)}(t)+l_k^{(01)}(t)+l_k^{(11)}(t)\right) \\
&=M_0^t.
\end{align*}
From the symmetry between $0$ and $1$, the same discussion applies to the number of $1$s.
Thus we have proved \eqref{01num_eq}. 
\qed\vspace*{5pt}

The following Corollary immediately follows from Proposition~\ref{sect2.2.proofprop1}.
\begin{Corollary}\label{sect2.2.proofCor1}
The number of $0$s and that of $1$s are conserved in time for a stationary state. 
\end{Corollary}

The following Proposition implies that the stationary solutions are all that we have listed in the previous subsection.  
\begin{Proposition}\label{sect2.2.proofprop2}
A stationary state which contains $0$ or $1$ satisfies one of the following equations.  
\begin{subequations}
\begin{align}
{}^\forall i,&\;\; u_i^tu_{i+1}^t=0 \label{p2eqa}\\
{}^\forall i,&\;\; (1-u_i^t)(1-u_{i+1}^t)=0 \label{p2eqb}
\end{align}
\end{subequations}
\end{Proposition}

The following Lemma is essential to prove Proposition~\ref{sect2.2.proofprop2}.
\begin{Lemma}\label{sect2.2.proofPr_lem1}
In a stationary state with $0$ or $1$, if there exists an element $\ast$ other than $0$ and $1$,  both of its two adjacent elements are $0$ or $1$, that is, $\ast$ exists in a pattern
 $0\ast 0$ or $1 \ast 1$.
\end{Lemma}
\Proof
In a stationary state, from the proof for Proposition~\ref{sect2.2.proofprop1}, we find that 
\begin{equation}\label{sect2.2.proof.bt}
b(t)=\sum_k l_k^{(00)}(t)+l_k^{(01)}(t),
\end{equation}
where $b(t)$ is the number of the boundaries between the patterns $P^{(ij)}$ in the form $0 \ast 0$ ($\ast \ne 0,\,1$）.
Equation \eqref{sect2.2.proof.bt} implies that all the patterns $P^{(00)}$ and $P^{(01)}$ consist the boundaries of that form.
Hence the leftmost $0$ of the patterns $P^{(00)}$ and $P^{(01)}$ must be the right $0$ of this boundary ($0\ast\underbar{0}$),
that means there is no sequence of the form $\DIS 0\underbrace{a_1 \cdots a_k}_{\mbox{\scriptsize $k\ge 1$}}\ast \,0$ ($a_i \ne 1$).

Similarly, from the symmetry between $0$ and $1$,  there is no sequence of the form $\DIS 1\,\ast \underbrace{b_1 \cdots b_k}_{\mbox{\scriptsize $k\ge 1$}}1$ ($b_i \ne 1$).
Thus the proof was completed. 
\qed
\vspace*{5pt}

\textbf{Proof of Proposition~\ref{sect2.2.proofprop2}}
\begin{itemize}
\item From Lemma~\ref{sect2.2.proofPr_lem1}, if there is no $1$, only the pattern $P^{(00)}$ among the four patterns exists and \eqref{p2eqa} holds. Similarly，if there is no $0$, \eqref{p2eqb} holds.
\item If there is no element other than $0$ and $1$, FCA 184 turns to the rule 184 CA, which has been investigated in detail as a traffic model\cite{Wolfram,Nishinari-Takahashi}.
When the number of $0$ is equal to or greater than the number of $1$, the state converges to a free-flow state where both of the two adjacent elements of $1$ are $0$.
Hence we have \eqref{p2eqa}.
Similarly,  if the number of $0$ is less than the number of $1$,  the state converges to a congestion state where both of the two adjacent elements of $0$ are $1$, 
and we have \eqref{p2eqb}.
\item When $0$, $1$, and $\ast (\ne 0,\,1)$ are mixed, from Lemma~\ref{sect2.2.proofPr_lem1},  we suppose that there are two elements $\gamma(\ne 0,1)$ and $\gamma'(\ne 0,1)$ at a time step in the form of $0 \gamma 0$ and $1 \gamma' 1$. 
From \eqref{sect2.2.proofeq1}, noticing the facts that 
$v_{n+1}^t=0$ implies $v_n^{t+1}=v_n^t$, and that $v_{n+1}^t=1$ implies $v_n^{t+1}=v_n^{t+2}$,
we find the time evolution rule gives 
\[
0\, \gamma \,0 \, \rightarrow \,  0\, \gamma \,a  ,\qquad
b\, 1\, \gamma'\, 1\,  b' \, \rightarrow \,  \gamma'\, 1\, b'\, c\, c',
\]
where $a$, $b$, $b'$, $c$, and $c'$ are some values depending on the state.
For a stationary state，Lemma~\ref{sect2.2.proofPr_lem1} implies that $a=0$ and that the left adjacent element to $\gamma'$ is equal to $1$.
Therefore a pattern $0\,\gamma \,0$ does not change its position and a pattern $1\,\gamma'\, 1$ moves two cites to the left.
Hence after a certain time steps, one of the following patterns is realized
\[
0 \, \gamma \, 0\, 1\, \gamma'\, 1,\qquad 0 \, \gamma \, 0\,  0\,1\, \gamma'\, 1,\qquad 0 \, \gamma \, 0\,  1\,1\, \gamma'\, 1.
\]
At the next time step, these patterns change to 
\[
0 \, \gamma \, \gamma'\, \cdots ,\qquad 0 \, \gamma \, 0\,  \gamma'\,1\, \cdots,\qquad 0 \, \gamma \, 1\,  \gamma'\,1\, \cdots 
\]
respectively, all of which cannot exist in a stationary state.
Thus we find that $0\ast 0$ and $1 \ast 1$ do not coexist in a stationary state.

Suppose that there exist patterns $0 \ast 0$.
Since a sequence $0\ast 0$ does not move in time, the time evolution of the other cells does not change by the transformation $\ast \rightarrow 1$.
Hence, by the consideration of the case where only $0$ and $1$ exist, a sequence $0\ast 0$  can exists when the number of $0$s is greater than or equal to the number of other values
in a stationary state, and \eqref{p2eqa} holds.
While a sequence $1 \ast 1$ exists, because of the symmetry between $0$ and $1$,  \eqref{p2eqb} holds.  
\end{itemize}
Thus Proposition~\ref{sect2.2.proofprop2} was proved.
\qed
\vspace*{10pt}

Since Proposition~\ref{sect2.2.proofprop2} indicates that either a free-flow solution or a anti-free flow solution is allowed in a stationary state which contains $0$ or $1$, and otherwise a stationary state is restricted to a uniform solution or a two-periodic solution, we obtain the following theorem.
\begin{Theorem}\label{sec2.2.theorem}
All the stationary solutions are the uniform solutions and the travelling wave solutions listed in the subsection~\ref{sect2_1}. 
\end{Theorem}

Finally let us discuss stability of the stationary solutions.
\begin{Theorem}\label{sect2.2.stability}
Free flow or anti-free flow solutions are unstable.
While two-periodic or uniform solutions are stable, but not asymptotically stable. 
\end{Theorem}
\Proof
From the Proposition~\ref{Prop_Betel}, a free flow solution will translate to a certain two-periodic or a uniform solution by small perturbation.
Hence a free flow solution is unstable and so is an anti-flow solution.
Let us consider a uniform solution $\rho_n^t=\alpha$.
Suppose that it is perturbed by small fluctuation as $\rho_n^0=\alpha+\delta_n$ at $t=0$.
Let $\epsilon:=\max_n[|\delta_n|]$.
By the same arguments in the proof of Lemma~\ref{sec2.2.lemma1}, we can show that $(\max_n[\rho_n^t])_{t=0}^\infty$ is a monotonically decreasing sequence in time $t$ and that $(\min_n[\rho_n^t])_{t=0}^\infty$ is a monotonically increasing sequence.
Thus it holds that $|\rho_n^t - \alpha| \le \epsilon$, which implies $\rho_n^t$ is stable in the sense of Lyapunov\cite{Bhatia-Szego}.
However, in general, it will not converge into the same uniform state and is not asymptotically stable.
In a similar way, a two-periodic solution is proved to be stable but not asymptotically stable.
\qed  
\begin{figure}[bht]
\centering
\begin{minipage}{0.5\hsize}
\begin{center}
\includegraphics[width=.8\textwidth]{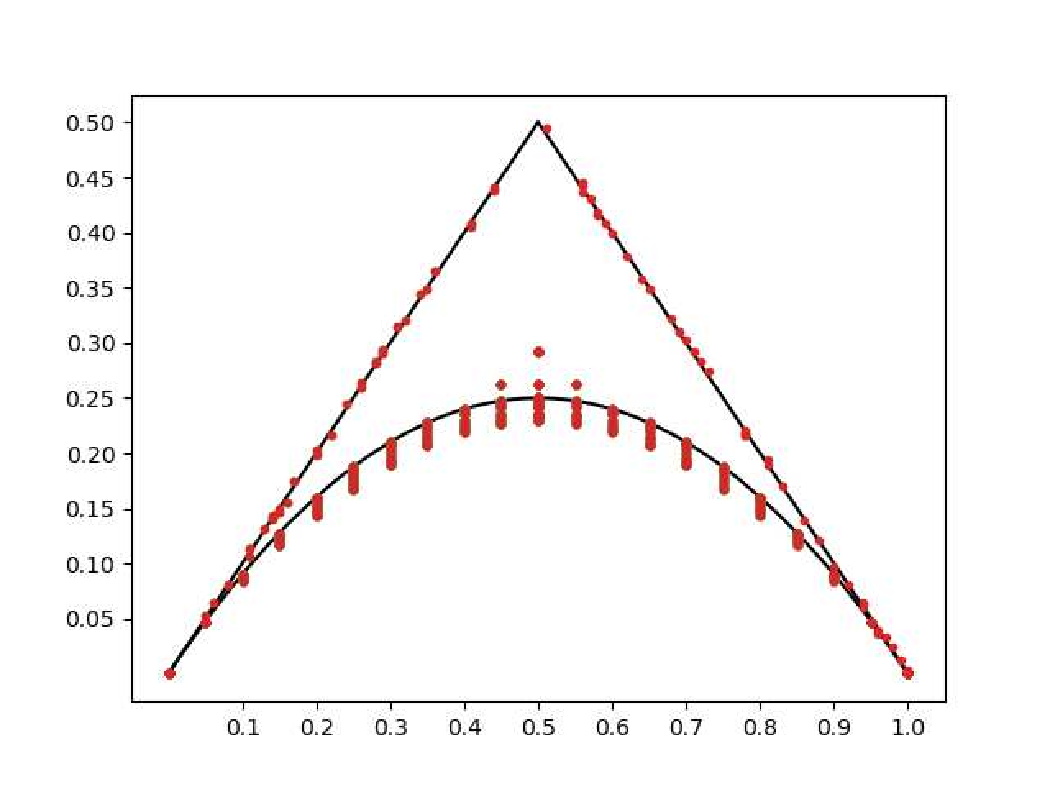}
\hspace{1.0cm}t=0
%\label{fig:1.5}       % Give a unique label
\end{center}
\end{minipage}%
\begin{minipage}{0.5\hsize}
\begin{center}
\includegraphics[width=.8\textwidth]{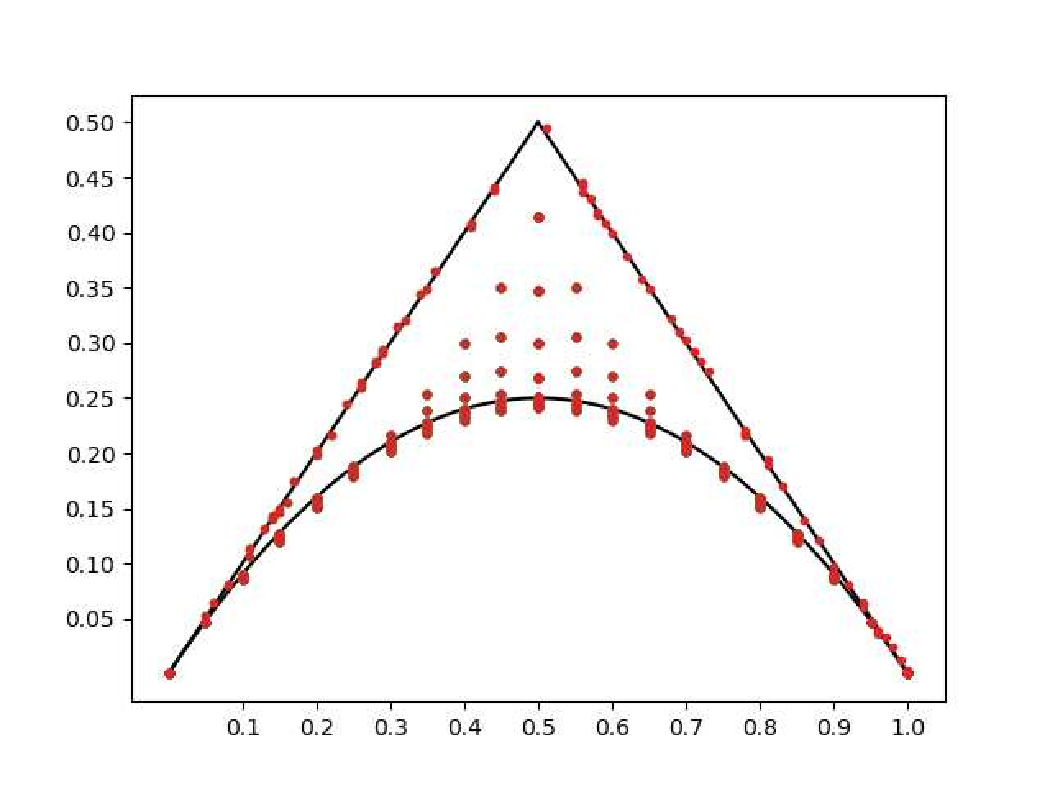}
\hspace{1.0cm}t=100
\end{center}%
\end{minipage}
\begin{minipage}{0.5\hsize}
\begin{center}
\includegraphics[width=.8\textwidth]{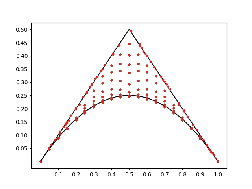}
\hspace{1.0cm}t=500
\end{center}
\end{minipage}%
\begin{minipage}{0.5\hsize}
\begin{center}
\includegraphics[width=.8\textwidth]{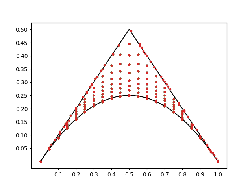}
\hspace{1.0cm}t=800
\end{center}%
\end{minipage}
\caption{The transient behaviour of the fundamental diagrams in case of even $N$ ($N=600$).
A flux increases in time and a state asymptotically turns to a two-periodic state.}
\label{fig:3}
\end{figure}
Figures~\ref{fig:3} shows an example of temporal change of the fundamental diagram for even $N$. 
Here we chose the initial state as
\begin{equation}
\left\{
\begin{array}{ll}
\rho^{t=0}_n = 2 \langle \rho \rangle \beta &\quad n = 1,2,...,\frac{N}{2}\\
\rho^{t=0}_n = 2 \langle \rho \rangle(1-\beta) &\quad n = \frac{N}{2}+1,...,N
\end{array}
\right.,
\end{equation}
where $ \langle \rho \rangle$ is the average density and $\beta$ ($0 \le \beta \le \frac{1}{2}$) is a parameter.
The system converges into various two-periodic states by changing $\beta$.
The flux always increases in time under this initial condition, though it can decrease in general.
\subsection{Fixed boundary conditions and bottle-neck effect}\label{sect2.3}

On highways, car density and/or car flux may differ place to place.
In particular, they discontinuously change at entrance and exit.
They also change at the place where the number of lanes decreases or increases.
With these situations in mind, let us consider FCA 184 with the boundary condition
\begin{equation}\label{sect2.3.boundary1}
\rho_0^t=a,\qquad \rho_{0}^t(1-\rho_1^t)=\lambda \qquad (0 < a \le 1,\; 0 \le \lambda \le a).
\end{equation}
The boundary condition \eqref{sect2.3.boundary1} may correspond to the case where the density and flux of cars at the entrance are controlled to be constant.
For a stationary state, there is neither a travelling wave solution nor a uniform solution except for $\lambda=a(1-a)$ due to the boundary condition \eqref{sect2.3.boundary1}.
However, some time-independent solutions may exist and we examine them.

In \eqref{sec1_FCA184}, we assume that  $\rho_n^t$ does not depend on $t$ and put $u_n=\rho_n^t$.
Then we have the three terms recurrence relation 
\begin{equation}\label{sec2.3.stationary}
u_{n+1}=\frac{u_n-u_{n-1}+u_nu_{n-1}}{u_n},\quad (u_0=a,\; u_0(1-u_1)=\lambda)
\end{equation} 
Equation~\eqref{sec2.3.stationary} has a conserved quantity
\[
u_{n+1}(1-u_n)=u_n(1-u_{n-1})=\lambda,
\]
and can be written as
\begin{equation}\label{sec2.3.u-lambda_eq}
u_{n+1}=1-\frac{\lambda}{u_n}.
\end{equation}
Thus $u_n$ can be obtained by continued fraction expansion as
\begin{equation}\label{sec2.3.continuedfrac}
u_n=U_n(\lambda;a):=1-\cfrac{\lambda}{1-\cfrac{\lambda}{\ddots\;\; 1- \cfrac{\ddots\;\;\lambda\quad}{1-\cfrac{\lambda}{a}}}}.
\end{equation}
Hence if and only if ${}^\forall n,\, 0 \le U_n(\lambda,a) \le 1$, there exists a time independent solution.  
\begin{figure}[bht]
\begin{minipage}{0.5\hsize}
\centering
\includegraphics[width=1.1\textwidth]{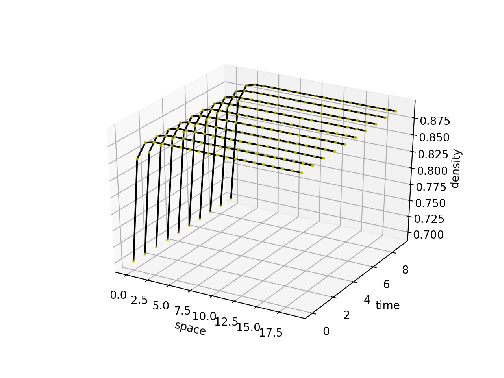}
\end{minipage}
\begin{minipage}{0.5\hsize}
\centering
\includegraphics[width=1.1\textwidth]{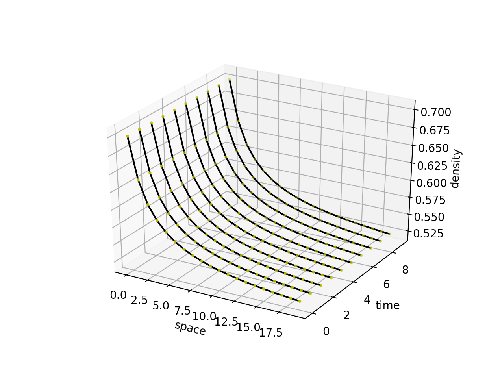}
\end{minipage}
\caption{Time independent solutions for the fixed boundary conditions. The parameters are $\lambda=0.1$,$a=0.7$ (left) and  $\lambda=0.25$,$a=0.7$ (right).}
\label{fig:6}       % Give a unique label
\end{figure}%
\begin{Proposition}\label{sec2.3.stationaryProp}
Time-independent solutions of \eqref{sec2.3.stationary} exist if and only if the following conditions are satisfied;
\begin{equation}\label{conditon}
\lambda \le \frac{1}{4} \quad \mbox{and} \quad \frac{1}{2}-\sqrt{\frac{1}{4}-\lambda} \le a. 
\end{equation}
\end{Proposition}
To prove Proposition~\ref{sec2.3.stationaryProp}, we use the following Lemma. 
\begin{Lemma}
\begin{equation}\label{sec2.3UnPoly}
U_n(\lambda;a)=\frac{(t_+^{n+1}-t_-^{n+1})a+(-t_+^{n+1}t_-+t_-^{n+1}t_+)}{
(t_+^{n}-t_-^{n})a+(-t_+^{n}t_-+t_-^{n}t_+)}, 
\end{equation}
where $t_{\pm}$ are the two roots of the algebraic equation $t^2-t+\lambda=0$:
\begin{equation}\label{sec2.3.tpm}
t_\pm(\lambda)=\frac{1 \pm \sqrt{1-4\lambda}}{2}.
\end{equation}
\end{Lemma}
\Proof
We define two polynomials of $\lambda$, $p_n(\lambda)$ and $q_n(\lambda)$, as
$p_0(\lambda)=a,\;\; q_0(\lambda)=1$, and 
\begin{subequations}
\begin{align}
p_{n+1}(\lambda)&=p_n(\lambda)-\lambda q_n(\lambda) \label{sec2.3.peq}\\
q_{n+1}(\lambda)&=p_n(\lambda) \label{sec2.3.qeq}.
\end{align}
\end{subequations}
Clearly, \eqref{sec2.3.u-lambda_eq} gives 
\begin{equation}\label{sec2.3.upq}
u_n=\frac{p_n(\lambda)}{q_n(\lambda)}.
\end{equation}
Since \eqref{sec2.3.peq}  and \eqref{sec2.3.qeq} are simultaneous linear recurrence relations,  we obtain 
\begin{align}
\begin{pmatrix}
p_n(\lambda)\\
q_n(\lambda)
\end{pmatrix}
&={\begin{pmatrix}
1&-\lambda \\
1&0
\end{pmatrix}
}^n
\begin{pmatrix}
a\\
1
\end{pmatrix}
\nonumber\\
&=\frac{1}{t_+-t_-}
\begin{pmatrix}
t_+&t_-\\
1&1
\end{pmatrix}
{\begin{pmatrix}
t_+&0\\
0&t_-
\end{pmatrix}}^n
\begin{pmatrix}
1&-t_-\\
-1&t_+
\end{pmatrix}
\begin{pmatrix}
a\\
1
\end{pmatrix}
\nonumber\\
&=\frac{1}{t_+-t_-}
\begin{pmatrix}
(t_+^{n+1}-t_-^{n+1})a+(-t_+^{n+1}t_-+t_-^{n+1}t_+) \\
(t_+^{n}-t_-^{n})a+(-t_+^{n}t_-+t_-^{n}t_+) 
\end{pmatrix},
\label{sec2.3.matform}
\end{align}
Thus we obtained \eqref{sec2.3UnPoly}.
\qed
\vspace*{5pt}
\textbf{Proof of Proposition~\ref{sec2.3.stationaryProp}} \par 
We give the proof in the three cases: (1) $\frac{1}{4}<\lambda \le 1$, (2) $\lambda=\frac{1}{4}$, and (3) $0\le \lambda <\frac{1}{4}$. \vspace*{5pt}
\\
(1) For $\frac{1}{4}<\lambda \le 1$, $t_\pm$ can be expressed as 
\[
t_\pm=\sqrt{\lambda}\,\e^{\sqrt{-1}\theta_\lambda}, 
\]
where $\theta_\lambda$ satisfies $\cos \theta_\lambda=\frac{1}{2\sqrt{\lambda}}$ and $\sin\theta_\lambda=\sqrt{1-\frac{1}{4\lambda}}$.
From Eq~\eqref{sec2.3UnPoly}, we obtain
\begin{align}
U_n(\lambda;a)&=\sqrt{\lambda}\,\frac{a\sin(n+1)\theta_\lambda -\sqrt{\lambda}\sin n \theta_\lambda}{a\sin n\theta_\lambda -\sqrt{\lambda}\sin (n-1) \theta_\lambda}
\nonumber \\
&=\sqrt{\lambda}\,\frac{\sin\left(n \theta_\lambda+\phi\right)}{\sin \left((n-1)\theta_\lambda+\phi\right)},
\end{align}
where 
\[
 \phi:=\tan^{-1}\left(\frac{a\sin \theta_\lambda}{a\cos \theta_\lambda -\sqrt{\lambda}}   \right).
\]
Since $0 < \theta_\lambda \le \frac{\pi}{3}$, there exists a positive integer $n_0$ such that 
\[
\sin \left((n_0-1)\theta_\lambda+\phi\right) \le 0 \quad \mbox{and} \quad 0< \sin\left(n_0 \theta_\lambda+\phi\right),
\]
that implies $U_{n_0}(\lambda;a)<0$ or $\infty$, and the condition ${}^\forall n,\, 0 \le u_n \le 1$ is not satisfied.
\hbreak
(2) For $\lambda=\frac{1}{4}$, we have
\[
U_n(\frac{1}{4};a)=\frac{1}{2}\frac{(a-\frac{1}{2})n+a}{(a-\frac{1}{2})(n-1)+a}.
\]
For $a\ge \frac{1}{2}$, $\frac{1}{2} < U_n(\frac{1}{4};a) \le 1$ holds for any $n$ and a time-independent solution exists.
While for $a<\frac{1}{2}$, there exists $n_0 \in \Z_{>0}$ such that $(a-\frac{1}{2})n_0+a \ge 0$ and $(a-\frac{1}{2})(n_0+1)+a \ge 0 <0$, 
that implies $U_{n_0+1}(\lambda;a)<0$ or $\infty$, and the condition ${}^\forall n,\, 0 \le u_n \le 1$ is not satisfied.
\hbreak
(3) When $\lambda<\frac{1}{2}$, we have
\[
U_1(\lambda;a)=1-\frac{\lambda}{a},\quad  U_n(\lambda;a)=T_n(\lambda)\frac{T_{n+1}(\lambda)a-\lambda}{T_n(\lambda)a-\lambda}\;\;(n \ge 2),
\] 
where
\[
T_n(\lambda):=\frac{t_+^n-t_-^n}{t_+^{n-1}-t_-^{n-1}}=t_+\,\frac{1-\left(\frac{t_-}{t_+}\right)^n}{1-\left(\frac{t_-}{t_+}\right)^{n-1}}.
\]
Let $\delta :=\frac{t_-}{t_+}$. Because 
\[
T_{n+1}(\lambda)-T_{n}(\lambda)=-\frac{\delta^{n-1}(1-\delta )^2}{(1-\delta^n )(  1-\delta^{n-1})}<0,
\]
$T_n(\lambda)$ is a monotonically decreasing function with respect to $n$, and
\[
T_2(\lambda)=1,\quad \lim_{n \to \infty}T_n(\lambda)=t_+.
\] 
Hence, if $t_+a-\lambda<0$, there exists $n_0$ such that $aT_{n_0}-\lambda \ge 0$ and $a T_{n_0+1}-\lambda<0$, 
that implies $U_{n_0+1}(\lambda;a)<0$ or $\infty$, and the condition ${}^\forall n,\, 0 \le u_n \le 1$ is not satisfied.
While,  if $t_+a-\lambda \ge 0$, 
\[
U_n(\lambda;a)<T_n(\lambda) \le 1
\]
and a time-independent solution exists. Since
$
t_+a-\lambda \ge 0 \; \Leftrightarrow  \; a \ge t_-,
$
the proof is completed.
\qed\vspace*{10pt}
%
%
%\begin{figure}[bht]
%\centering
%\includegraphics[width=.5\textwidth]{./fig/op(0.4to0.9).eps}
%\caption{bottleneck effect in the where $u^t_n$ takes from 0.4 to 0.9}
%\label{fig:8}       % Give a unique label
%\end{figure}%
%
%
\begin{figure}[bht]
\begin{minipage}{0.5\hsize}
\centering
\includegraphics[width=.9\textwidth]{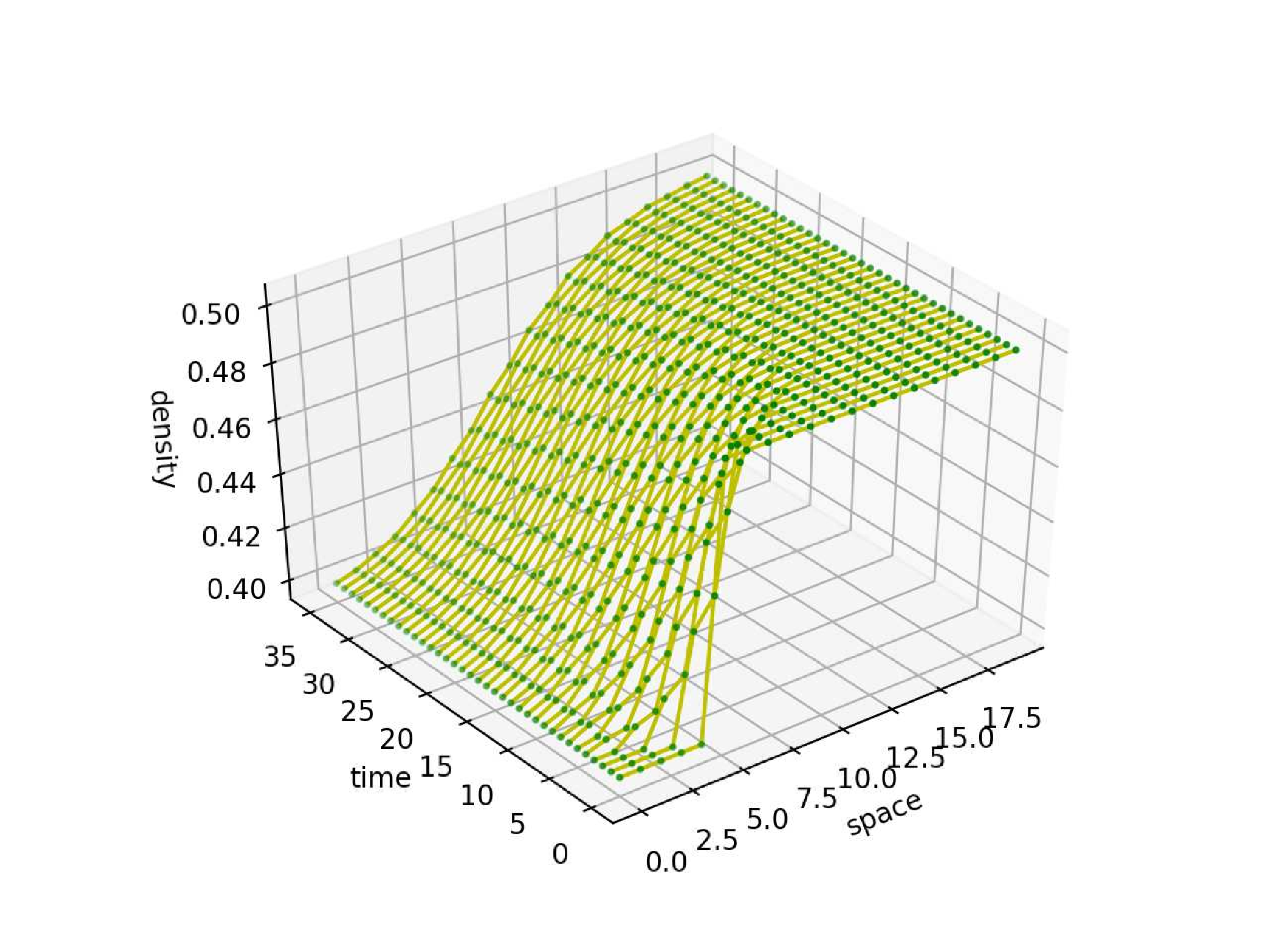}
\end{minipage}
\begin{minipage}{0.5\hsize}
\centering
\includegraphics[width=.9\textwidth]{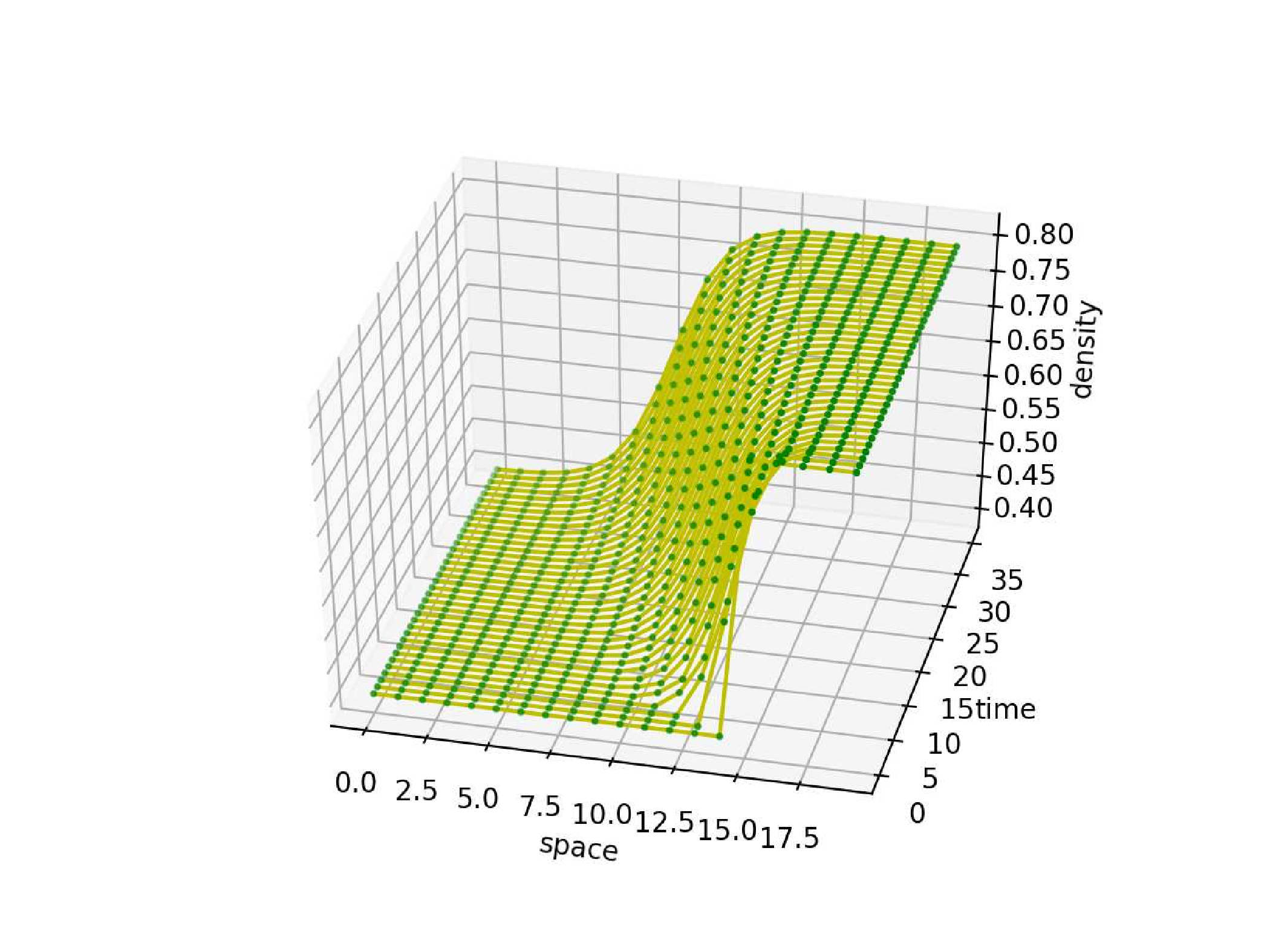}
\end{minipage}
\caption{Bottle-neck effect of traffic flow is shown. The congested regions propagate to the backward direction.
The car density $u^t_n$ changes from $0.4$ to $0.8$ (left), and from $0.4$ to $0.5$ (right).}
\label{fig:9}       
\end{figure}%

If traffic congestion takes place, the car flux at the entrance of a highway is different from that at the exit.
Since a car flux is constant in a  time-independent state, the congestion is a transient phenomenon.
Figures~\ref{fig:9} show the time evolution of FCA 184 with the boundary condition where the flux at the entrance is larger than that at the exit.
We find that the high density region of cars is extending  backward as is empirically seen in traffic jams.
To examine this kind of transient behaviour analytically, we consider ultradiscrete limit of FCA 184 in the next section.
%
%  
%%%%%%%%%%%%%%%%%%%%%%%%%%%%%%%%%%%%%%%
%
%

%%%%%%%%%%%%%%%%%%%%%%%%%%%%%%%%%%%%%%
\section{Ultradiscrete analysis for FCA 184}
\label{sec4.ultra}
Ultradiscretisation is a limiting procedure which transforms a given equation to a piece-wise linear equation\cite{Tokihiro_etal}.
For FCA 184, we put
\begin{equation}\label{3.ultra_rho}
\rho_n^t=\e^{-U_n^t/\epsilon},
\end{equation}
and construct equations for $U_n^t$ by taking $\epsilon \rightarrow +0$.
However, since  \eqref{sec1_FCA184} contains a negative term, it is not
straightforward to take a limit.
A method to avoid this difficulty is to use the method of ultradiscretisation with sign\cite{Mimura_etal},
but here we use another approach.
Let $q_n^t:=1-\rho_n^t$. 
Then, from \eqref{sect2_eq1_r}, we find 
\begin{subequations}
\begin{align}
\rho_n^{t+1}&=\rho_{n-1}^tq_n^t+\rho_n^t\rho_{n+1}^t \label{sec3.eq1}\\
q_n^{t+1}&=q_{n+1}^t\rho_n^t+q_n^tq_{n-1}^t \label{sec3.eq2}\\
1&=\rho_n^t+q_n^t \label{sec3.RQconserve}.
\end{align}
\end{subequations}
Introducing $V_n^t$ by
\begin{equation}\label{sec3.ultraV}
q_n^t=\e^{-V_n^t/\epsilon},
\end{equation}
and taking the limit $\epsilon \rightarrow +0$, we obtain the following simultaneous equations. 
\begin{subequations}
\begin{align}
U_n^{t+1}&=\min\left[ U_{n-1}^t+V_n^t, U_n^t+U_{n+1}^t \right]\label{sec3.Ueq1}\\
V_n^{t+1}&=\min\left[ V_{n+1}^t+U_n^t,V _n^t+V_{n-1}^t \right]\label{sec3.Ueq2}\\
0&=\min\left[ U_n^t, V_n^t \right]\label{sec3.URQconserve}.
\end{align}
\end{subequations}
Note that $U_n^t,\,V_n^t \in \R_{\ge 0} \sqcup \{\infty\}$ due to the inequality $0 \le \rho_n^t, q_n^t \le 1$. 
\begin{Proposition}\label{sec3.propUV0}
If ${}^\forall n,\,\min[U_n^0,V_n^0]=0$, then ${}^\forall n,\,{}^\forall t,\, \min[U_n^t,V_n^t]=0.$
\end{Proposition}
\Proof 
We prove by induction.

For $t=0$, the statement is trivial.
Suppose that it holds for $t=k$.
If $U_n^{k+1} \ne 0$, one of the pair $\{U_{n-1}^k, \, V_n^k\}$ and one of $\{U_n^k,\, U_{n+1}^k\}$ are not equal to $0$.
Since one of $\{U_n^k,\,V_n^k\}$ is equal to $0$, non-zero pairs must be $\{U_{n-1}^k,\, U_n^k\}$, $\{U_{n-1}^k,\,U_{n+1}^k\}$, or $\{V_n^k,\,U_{n+1}^k\}$,
that means one of the following three pairs are both zero $\{V_{n-1}^k, \, V_n^k\}$, $\{V_{n-1}^k, \, V_{n+1}^k\}$ and $\{U_n^k,\, V_{n+1}^k\}$.
The first pair and the third pair give $V_n^{k+1}=0$, and the second pair also gives $V_n^{k+1}=0$ because one of $U_n^k$ and $V_n^k$ is zero.
Thus the Proposition is true for $t=k+1$.
From the induction hypothesis, the Proposition is true for any $t$, which completes the proof.
\qed
\vspace*{5pt} 

Proposition~\ref{sec3.propUV0} means that once the initial state
satisfy \eqref{sec3.URQconserve}, it is satisfied forever, and the dynamical system described by \eqref{sec3.Ueq1} and \eqref{sec3.Ueq1} is deterministic.
If ${}^\forall n,\, U_n^0, V_n^0 \in \{0, \infty\}$, the dynamical system is equivalent to the rule 184 CA because $0$ and $\infty$ correspond to $1$ and $0$ in the rule 184 CA.

Firstly we examine transient behaviour in the region $0<\rho_n^t \ll 1$ (free-flow region).
In the ultradiscrete limit, this situation will correspond to the case ${}^\forall n,\, V_n^0=0$.
We also assume that the initial state satisfies the following condition:
\begin{equation}\label{sec3.Uboundaryab}
U_n^0>0,\quad \mbox{and}\quad U_n^0=\alpha \;\;( n \ge N), \quad U_n^0=\beta \;\; ( n \le -N),
\end{equation}
where $N$ is a positive integer.
Note that, due to the transformation $U_n^t=-\lim_{\epsilon \to +0} \log \rho_n^t$, the smaller the value $U_n^t$, the higher the density $\rho_n^t$.

\begin{Proposition}\label{sec3.ud.prop}
Any state given by \eqref{sec3.Ueq1} -- \eqref{sec3.URQconserve}, that satisfies ${}^\forall n,\, V_n^0=0$ and \eqref{sec3.Uboundaryab}, 
turns to be a stationary travelling wave state with velocity one in finite time steps.
\end{Proposition}

To prove Proposition~\ref{sec3.ud.prop}, we prepare several notations and Lemmas.
We put $U_n^t=:X_{n-t}^t$, then, 
from Proposition~\ref{sec3.propUV0}, \eqref{sec3.propUV0} becomes
\begin{equation}\label{sec3.U_eq}
X_n^{t+1}=\min\left[ X_{n}^t, X_{n+1}^t+X_{n+2}^t\right],
\end{equation}
with boundary condition
\begin{equation}\label{sec3.X.boundary}
X_n^0>0,\quad \mbox{and}\quad X_n^0=\alpha \;\;( n \ge N), \quad X_n^0=\beta \;\; ( n \le -N).
\end{equation}
The following Lemmas are readily proved by induction with respect to $t$. 
\begin{Lemma}\label{sec3.Lem.decreas}
\begin{equation}\label{sec3.eq_X_ineq}
{}^\forall t,\, {}^\forall n,\; X_n^{t+1} \le X_n^t 
\end{equation}
\end{Lemma}
\begin{Lemma}\label{sec3.Lem.two.const}
If there exists a time step $t_0$ such that $X_{n+1}^t=X_{n+1}^{t_0}$ and $X_{n+2}^t=X_{n+2}^{t_0}$ for ${}^\forall \, t \ge t_0$.
Then, ${}^\forall\, t \ge t_0+1$, $X_n^t=X_n^{t_0+1}$.
\end{Lemma}
\begin{Lemma}\label{sec3.Lem_ineq}
If $Y_n^t$ is a solution of \eqref{sec3.U_eq} with the initial condition that satisfies ${}^\forall n,\, X_n^0 \ge Y_n^0$, then,
it holds that ${}^\forall n,\,{}^\forall t,\, X_n^t \ge Y_n^t$.
\end{Lemma}

Let $\gamma$ be the minimum value among $\{ \beta, X_{-N+1}, X_{-N+2}, \ldots, X_{N-1}, \alpha\}$, and let $Y_n^t$ be the solution of  \eqref{sec3.U_eq}
with the boundary condition
\begin{equation}\label{sec3.initial_Y}
Y_N^0=\left\{
\begin{array}{cl}
\gamma&\ (-N+1 \le n ) \\
\beta&(n \le -N)
\end{array}
\right.
.
\end{equation}
We also define the Fibonacci numbers $F_j$ $(j=0,1,2,...)$ as
\begin{Definition}\label{Def.Fibonacci}
\[
F_0=1,\; F_1=1,\;\; F_{j+1}=F_j+F_{j-1}\;\; (j=1,2,...).
\] 
\end{Definition}
For example, $F_2=2$, $F_3=3$, $F_4=5$ and $F_5=8$.

\begin{Lemma}\label{sect3.lem.Y_const}
Let $k_\gamma$ be the positive integer which satisfies
\begin{equation}\label{eq_prop_2}
F_{k_\gamma} \gamma \le \beta < F_{k_\gamma+1} \gamma.
\end{equation}
Then, for $n \le -N-k_\gamma+1$, $Y_n^t$ is constant in time, that is, 
\begin{equation}
Y_n^t=Y_n^0=\beta  \quad  (n \le -N-k_\gamma+1)
\end{equation}
\end{Lemma}
\Proof
From \eqref{sec3.U_eq} and \eqref{sec3.initial_Y}, we have
\[
Y_n^1=\left\{
\begin{array}{cl}
\gamma &\; (n \ge -N+1)\\
\min[\beta, 2\gamma] & \;(n=-N)\\
\beta &\; (-N-1 \le n)
\end{array}.
\right.
\]
In the next time step,
\[
Y_n^2=\left\{
\begin{array}{cl}
\gamma &\; (n \ge -N+1)\\
\min[\beta, 2\gamma] & \;(n=-N)\\
\min[\beta, Y_{-N}^1+\gamma] &\; (n=-N-1)\\
\beta &\; (-N-2 \le n)
\end{array}.
\right.
\]
By repeating this procedure，in the case $F_4\gamma \le \beta < F_5\gamma$ for example, we find the following
time evolution pattern
\[
\begin{array}{lcccccccccc}
  n &         &-N-4&-N-3&-N-2&-N-1&-N&-N+1&-N+2&  \\
t=0&\cdots&\beta&\beta&\beta&\beta&\beta&\gamma&\gamma&\cdots\\
t=1&\cdots&\beta&\beta&\beta&\beta&2\gamma&\gamma&\gamma&\cdots\\
t=2&\cdots&\beta&\beta&\beta&3\gamma&2\gamma&\gamma&\gamma&\cdots\\
t=3&\cdots&\beta&\beta&5\gamma&3\gamma&2\gamma&\gamma&\gamma&\cdots\\
t=4&\cdots&\beta&\beta&5\gamma&3\gamma&2\gamma&\gamma&\gamma&\cdots\\
t=5&\cdots&\beta&\beta&5\gamma&3\gamma&2\gamma&\gamma&\gamma&\cdots\\
%t=6&\cdots&\beta&\beta&5s&3s&2s&s&s&\cdots\\
\end{array}
\]
Thus, for $n=-N-i$ ($i=0,1,2$),  $Y_n^t$ becomes $F_{i+2}\, \gamma$ at $t=i+1$, and does not change afterwards,
and $Y_n^t$ is constant in time for $n \le -N-3$.

Let us consider general cases. 
Note that $Y_n^t=Y_n^0=\gamma$ for $-N+1 \le n$.

When $k_\gamma=1$，$Y_{-N}^1=\min[\beta, F_2\,\gamma]=\beta$ and  ${}^\forall n,\, X_n^1=X_n^0$. Hence, ${}^\forall\, t,\, {}^\forall \, n$, and Lemma~\ref{sect3.lem.Y_const} holds.

When $k_\gamma \ge 2$, $Y_{-N}^1=F_2\, \gamma$ and  $Y_n^1=\beta $ for $n \le -N-1$.
Since ${}^\forall t\,$, $Y_n^t=\gamma$ for $-N+1 \le n$,  $Y_{-N}^t=2\,\gamma$ for $t \ge 1$.

Similarly, for $t=i+1$ ($i=0,1,...,k_\gamma-2$),  only at the site $n=-N-i$, $Y_n^t$ changes its value as $Y_{-N-i}^{i}=\beta$ $\rightarrow$ $Y_{-N-i}^{i+1}=F_{i+2}\, \gamma$.

At $t=k_\gamma$,
\begin{align*}
Y_{-N-k_\gamma+1}^{k_\gamma}
&=\min[Y_{-N-k_\gamma+1}^{k_\gamma-1}, Y_{-N-k_\gamma+2}^{k_\gamma-1}+Y_{-N-k_\gamma+3}^{k_\gamma-1}]\\
&=\min[\beta, F_{k_\gamma}\, \gamma+F_{k_\gamma-1}\, \gamma]\\
&=\min[\beta, F_{k_\gamma+1}\,\gamma]=\beta=Y_{-N-k_\gamma+1}^{k_\gamma-1}.
\end{align*}
Thus we find that ${}^\forall n,\, Y_n^{k_\gamma}=Y_n^{k_\gamma-1}$, which implies that ${}^\forall n,\, Y_n^t=Y_n^{k_\gamma-1}$ for $t \ge k_\gamma$.
Therefore it holds that $Y_n^t=\beta$ for $n \le -N-k_\gamma+1$, which completes the proof.
\qed
\vspace*{10pt}

\textbf{Proof of Proposition~\ref{sec3.ud.prop}}\par
From Lemma~\ref{sec3.Lem.two.const} and ${}^\forall t,\, X_{N}^t=X_{N+1}^t=\alpha$, we have $X_{N-1}^t=X_{N-1}^1$ ($t \ge 1)$, $X_{N-2}^t=X_{N-2}^2$ $(t \ge 2)$, $\ldots$.
Hence for $t \ge j$, $X_n^t$ does not change if $n \ge N-j$.
On the other hand, from Lemma~\ref{sect3.lem.Y_const} and Lemma \ref{sec3.Lem_ineq}, it holds that 
\[
{}^\forall t,\, X_n^t \ge Y_n^t =\beta \quad (n \le -N-k_\gamma+1).
\]
However, from Lemma~\ref{sec3.Lem.decreas}, $X_n^t \le X_n^0 = \beta $ for $n \le -N$, and we have
\[
{}^\forall t,\, X_n^t=X_n^0=\beta \quad (n \le -N-k_\gamma+1).
\]
Therefore, for $t \ge 2N+k_\gamma-1$, ${}^\forall n,\,$ $X_n^t$ does not change in time, which completes the proof of Proposition~\ref{sec3.ud.prop}.
\qed
\vspace*{5pt}
Because of the symmetry between $U_n^t$ and $V_n^t$, we immediately have the following Corollary.
\begin{Corollary}\label{sec3.Cor1}
Any state given by \eqref{sec3.Ueq1}-- \eqref{sec3.URQconserve}, that satisfies ${}^\forall n,\, U_n^0=0$ and the boundary condition: 
\[
V_n^0>0,\quad \mbox{and}\quad V_n^0=\alpha \;\;( n \ge N), \quad V_n^0=\beta \;\; ( n \le -N),
\]
turns to be a stationary travelling wave state in finite time steps which moves to the opposite direction with velocity one. 
\end{Corollary}
The condition ${}^\forall n,\, U_n^0=0$ corresponds to a congestion limit, and Corollary~\ref{sec3.Cor1} means that the congestion wave propagates backward, that coincides with our empirical understanding.  
We also note that, if we consider a cyclic boundary condition with period $N$, then we can similarly prove that, in time steps less than $N$, any state turns to be a forward going travelling wave state with velocity one if ${}^\forall n,\, V_n^0=0$, and a backward going travelling wave state with velocity one if ${}^\forall n,\, U_n^0=0$.

\begin{figure}[bht]
\centering
\begin{minipage}{0.5\hsize}
\begin{center}
\includegraphics[width=.9\textwidth]{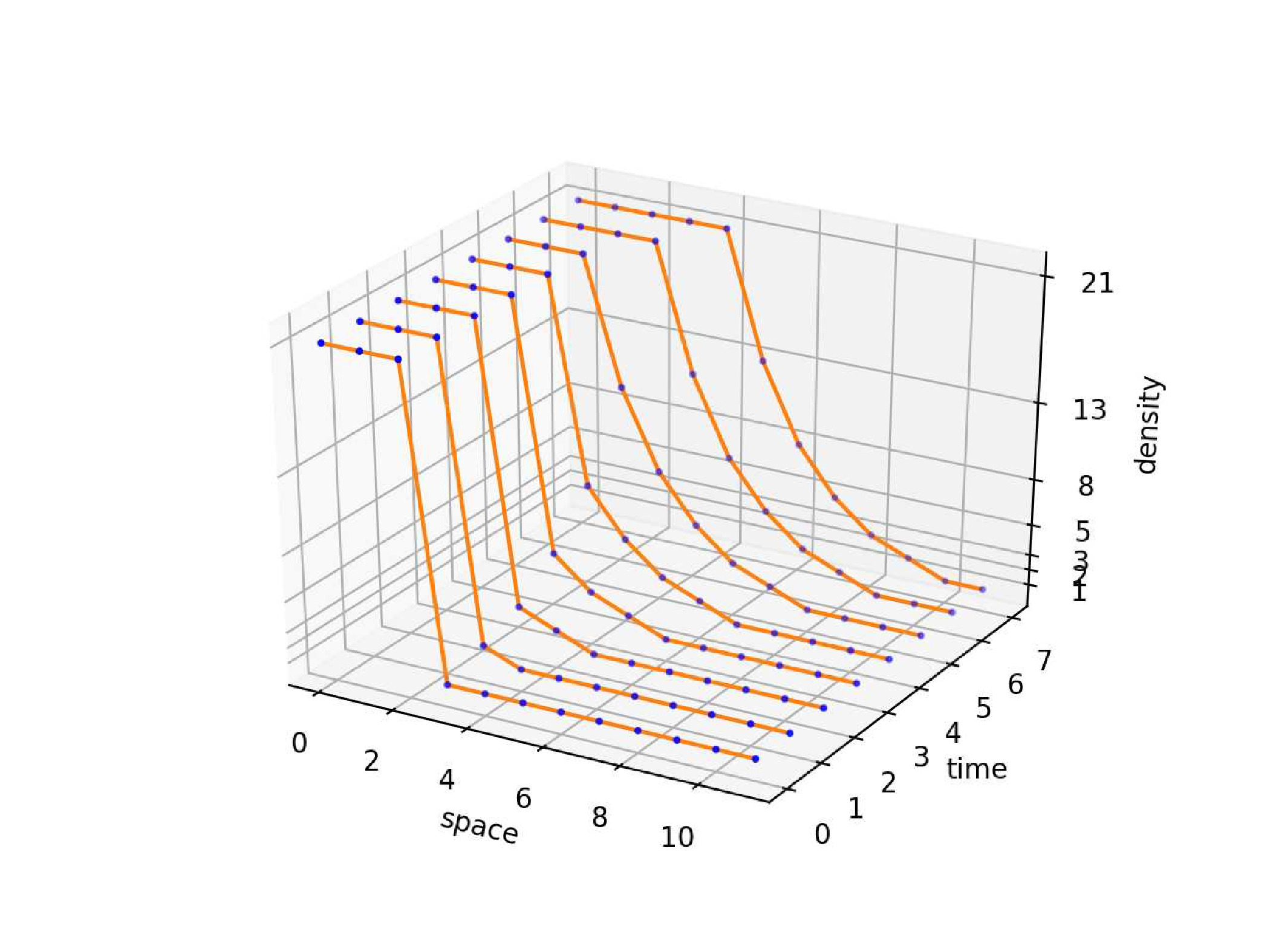}
\hspace{1.0cm}A Fibnacci type
\end{center}
\end{minipage}%
\begin{minipage}{0.5\hsize}
\begin{center}
\includegraphics[width=.9\textwidth]{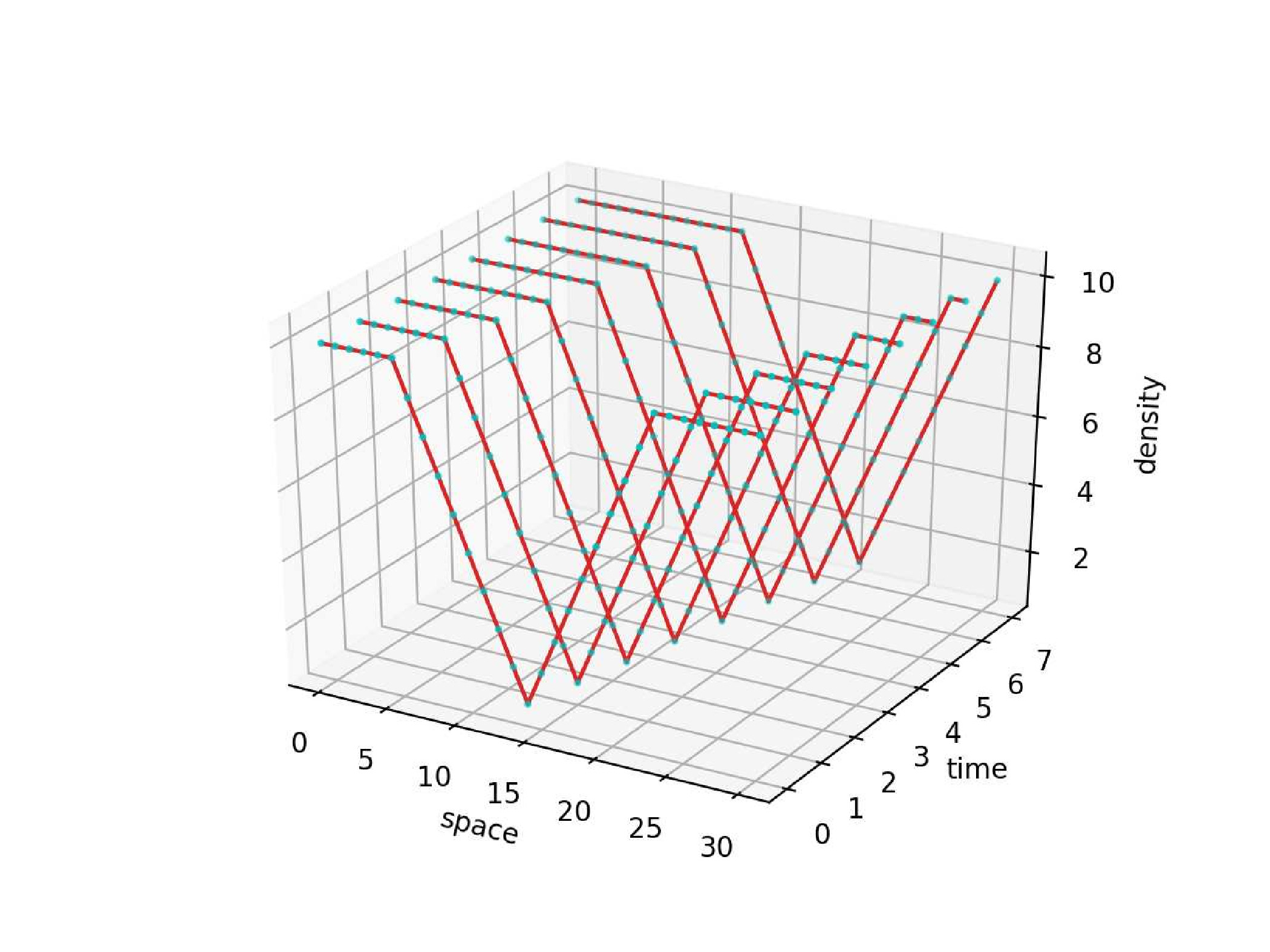}
\hspace{1.0cm}A triangle type
\end{center}%
\end{minipage}
%
%\begin{minipage}{0.5\hsize}
%\begin{center}
%\includegraphics[width=.9\textwidth]{UDv.eps}
%\hspace{1.0cm}A behavior of V
%\end{center}
%\end{minipage}%
%
\caption{Examples of time evolution of $U_n^t$ in the solutions of  \eqref{sec3.Ueq1}-\eqref{sec3.URQconserve}.}
\label{fig:11}
\end{figure}

Figures~\ref{fig:11} shows two examples of time evolution patterns.
The Fibonacci type solutions are obtained for the initial conditions given by step functions.  
For example, if $U_n^0$ is given for $k \ge 2$ as
\begin{equation}
U_n^0=\left\{
\begin{array}{cl}
F_k & (n\le 0)\\
1 & (n\ge 1)
\end{array}
\right.,
\end{equation}
then, after $k-2$ time steps, $U_n^t$ converges to the travelling wave state with velocity one as
\begin{equation}
U_n^t=\left\{
\begin{array}{cl}
F_k & (n-t\le k-2)\\
F_{k-1} & (n-t =k-1)\\
F_{k-2} & (n-t =k)\\
\cdots & \\
F_2 & (n-t=0)\\
1 & (n-t\ge 1)
\end{array}
\right.
\end{equation}
A triangle type solution also exhibits a travelling wave with velocity one.
For a positive integer $k$ ($k \ge 2$), it is given as
\begin{equation}
U_n^t=\left\{
\begin{array}{cl}
k & (n-t \le -k+1)\\
k-1 & (n-t = -k+2)\\
\cdots & \\
1 & (n-t=0)\\
2 & (n-t=1) \\
\cdots & \\
k & (n-t \ge k-1)
\end{array}
\right.
\end{equation}
We can construct various kinds of exact solutions by mixing these solutions.
It is also fairly simple to solve an initial value problem of the ultradiscrete equations \eqref{sec3.Ueq1}-\eqref{sec3.URQconserve}.
\section{Concluding remarks}
In this article, we investigated FCA184 as a mathematical model of traffic flow.
It includes the rule 184 ECA and the Burgers equation as special cases.
We obtained all the stationary solutions for the periodic boundary conditions  and proved their stability.
An interesting feature of this model is that, when the number of total sites is even, the fundamental diagram of stationary states is a two-dimensional domain and 
any point in the domain denotes a stable state.
We presume that this domain corresponds to the synchronized modes of traffic flow.
The bottle-neck effect was demonstrated by fixed boundary conditions and we gave the condition of the boundary values for the existence of time independent solutions.
The ultradiscrete limit of FCA184 was also discussed and proved that any state turns to a travelling wave state in finite time steps for generic initial conditions.
Extension of the FCA184 to more realistic models such as a slow start model is one of the problems we wish to address in the future. 
\begin{figure}[bht]
\centering
\includegraphics[width=.5\textwidth]{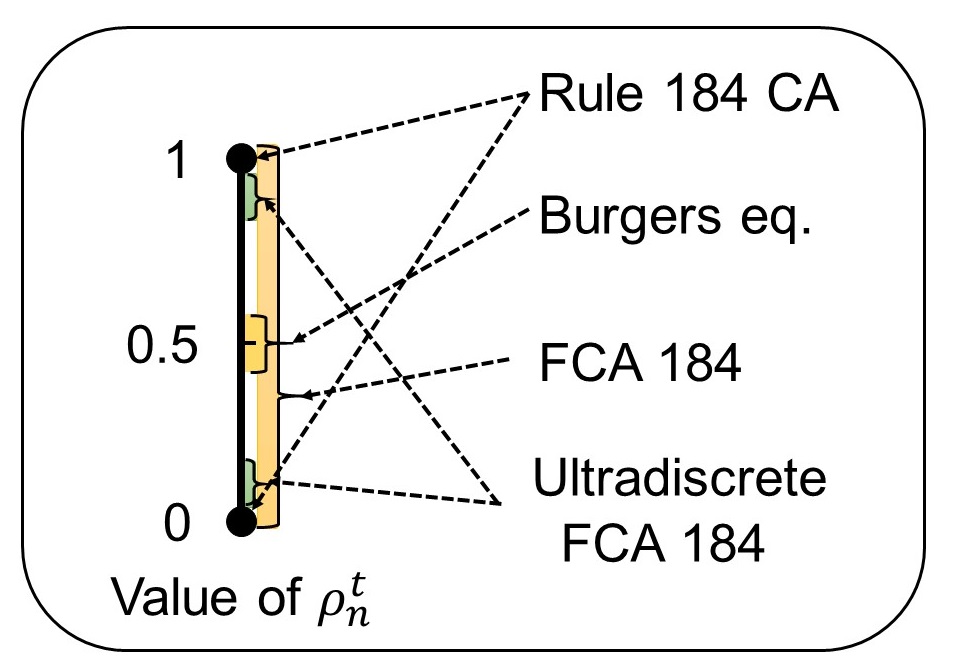}
\caption{Relation between FCA184 and other models. When we restrict the values $0$ and $1$, FCA184 becomes the rule 184 CA.
The Burgers equation is obtained by continuous limit aroung $\rho_n^t=0.5$. The ultradiscrete limit of FCA184 gives low and high density limit of FCA184.}
\label{fig:12}       % Give a unique label
\end{figure}

%\paragraph{Paragraph headings} Use paragraph headings as needed.

%\begin{figure}
%  \includegraphics{example.eps}
%\end{figure}
%\begin{figure*}
%  \includegraphics[width=0.75\textwidth]{example.eps}
%\end{figure*}
%\begin{table}
%\begin{tabular}{lll}
%\hline\noalign{\smallskip}
%\end{tabular}
%\end{table}

\begin{acknowledgements}
We would like to thank Prof. Ralph Willox and Dr. Takafumi Mase for useful comments.
KH expresses his sincere thanks to ADK Marketing Solutions Inc. for their scholarship support.
This work is supported in part by joint research funding from Arithmer Inc.
\end{acknowledgements}


\bibliographystyle{spmpsci}      % mathematics and physical sciences
\bibliography{FCA184_20200328}   % name your BibTeX data base

\begin{thebibliography}{10}
\providecommand{\url}[1]{{#1}}
\providecommand{\urlprefix}{URL }
\expandafter\ifx\csname urlstyle\endcsname\relax
  \providecommand{\doi}[1]{DOI~\discretionary{}{}{}#1}\else
  \providecommand{\doi}{DOI~\discretionary{}{}{}\begingroup
  \urlstyle{rm}\Url}\fi

\bibitem{Bando_etal}
Bando, M., Hasebe, K., Nakayama, A., Shibata, A., Sugiyama, Y.: Dynamical model
  of traffic congestion and numerical simulation.
\newblock Phys. Rev. E \textbf{51}, 1035--1042 (1995)

\bibitem{Betel-Flocchini}
Betel, H., Flocchini, P.: On the asymptotic behavior of fuzzy cellular
  automata.
\newblock Electronic Notes in Theoretical Computer Science \textbf{252}, 23--40
  (2009)

\bibitem{Bhatia-Szego}
Bhatia, N.P., {Szeg\"{o}}, G.P.: Stability Theory of Dynamical Systems.
\newblock Classics in Mathematics. Springer-Verlag, Berlin Heidelberg (1970)

\bibitem{Cattaneo_etal}
Cattaneo, G., Flocchini, P., Mauri, G., Quaranta-Vogliotti, C., Santoro, N.:
  Cellular automata in fuzzy backgrounds.
\newblock Physica D \textbf{105}, 105--120 (1997)

\bibitem{Elefteriadou}
Elefteriadou, L.: An Introduction to Traffic Flow Theory.
\newblock Springer Optimization and Its Applications 84. Springer
  Science+Buisiness Media, New York (2014)

\bibitem{Greenshield}
Greenshields, B.D.: A study of traffic capacity.
\newblock Proceedings of the Highway Research Board \textbf{14}, 448--477
  (1935)

\bibitem{Higashi_etal}
Higashi, K., Nukaya, T., Satsuma, J., Tomoeda, A.: On a new nonlinear discrete
  model describing traffic flow.
\newblock The bulletin of Musashino University Musashino Center of Mathematical
  Engineering \textbf{4}, 42--49 (2019)

\bibitem{Kerner-Rehborn}
Kerner, B.S., Rehborn, H.: Experimental properties of complexity in traffic
  flow.
\newblock Phys. Rev. E \textbf{53}, R4275--R4278 (1996)

\bibitem{Lighthill-Whitham}
Lighthill, M., Whitham, G.: On kinematical waves ii. a theory of traffic flow
  on long crowded roads.
\newblock Proc. R. Soc. London A \textbf{229}, 317--345 (1955)

\bibitem{Lubashevski_etal}
Lubashevsky, I., Mahnke, R., Wagner, P., Kalenkov, S.: Long-lived states in
  synchronized traffic flow: Empirical prompt and dynamical trap model.
\newblock Phys. Rev. E \textbf{66}, 016117--1 (2002)

\bibitem{Mimura_etal}
Mimura, N., Isojima, S., Murata, M., Satsuma, J.: Singularity confinement test
  for ultradiscrete equations with parity variables.
\newblock J. Phys. A: Math. Theor. \textbf{42}, 315206 (2009)

\bibitem{Musha-Higuchi}
Musha, T., Higuchi, H.: Traffic current fluctuation and the burgers equation.
\newblock Jpn. J. Appl. Phys. \textbf{17}, 811--816 (1978)

\bibitem{Nagatani}
Nagatani, T.: The physics of traffic jams.
\newblock Rep. Prog. Phys. \textbf{65}, 1331--1386 (2002)

\bibitem{Nagel-Schreckenberg}
Nagel, K., Schreckenberg, M.: A cellular automatonmodel for freeway traffic.
\newblock J. Phys.I France 2 \textbf{12}, 2221--2229 (1992)

\bibitem{Ni}
Ni, D.: Traffic Flow Theory: Characteristics, Experimental Methods, and
  Numerical Techniques.
\newblock Elsevier, New York (2015).
\newblock See Section 4-1 \textit{Single-Regime Models}

\bibitem{Nishinari-Takahashi}
Nishinari, K., Takahashi, D.: Analytic properties of ultradiscrete burgers
  equation and rule-184 cellular automaton.
\newblock J. Phys. A: Math. Gen. \textbf{31}, 5439--5450 (1998)

\bibitem{Pipes}
Pipes, L.A.: An operational analysis of traffic dynamics.
\newblock Journal of Applied Physics \textbf{24}, 274--281 (1952)

\bibitem{Sugiyama}
Sugiyama, Y., Yamada, H.: Simple and exactly solvable model for queue dynamics.
\newblock Phys. Rev. E \textbf{55}, 7749--7752 (1997)

\bibitem{Tokihiro_etal}
Tokihiro, T., Takahashi, D., Matsukidaira, J., Satsuma, J.: From soliton
  equations to integrable cellular automata through a limiting procedure.
\newblock Phys. Rev. Lett. \textbf{76}, 3247--3250 (1996)

\bibitem{Tomer_etal}
Tomer, E., Safonov, L., Havlin, S.: Presence of many stable nonhomogeneous
  states in an inertial car-following model.
\newblock Phys. Rev. Lett. \textbf{84}, 382--385 (2000)

\bibitem{Wolfram}
Wolfram, S.: Theory and Applications of Cellular Automata.
\newblock World Scientific, Singapore (1986)

\end{thebibliography}

%\begin{thebibliography}{}
% etc
%\end{thebibliography}

\end{document}